\documentclass[twocolumn,showpacs,preprintnumbers,nofootinbib,prd,
superscriptaddress,10pt]{revtex4-1}

\usepackage{amsmath,amssymb}
\usepackage[normalem]{ulem}
\usepackage{textcomp}
\usepackage{hyperref}
\usepackage{bm}
\usepackage{graphicx}
\usepackage{psfrag}
\usepackage[usenames,dvipsnames]{xcolor}
\usepackage[utf8]{inputenc}

\allowdisplaybreaks[4]

\newcommand{\be}{\begin{equation}}
\newcommand{\ee}{\end{equation}}
\newcommand{\bea}{\begin{eqnarray}}
\newcommand{\eea}{\end{eqnarray}}

\newcommand{\bs}{\begin{subequations}}
\newcommand{\es}{\end{subequations}}

\graphicspath{{./figures/}}

\begin{document}

\title{Solving post-Newtonian accurate Kepler Equation}

\date{\today}

\author{Yannick Boetzel}
\affiliation{Physik-Institut, Universit\"at Z\"urich,
Winterthurerstrasse 190, 8057 Z\"urich}

\author{Abhimanyu Susobhanan}
\affiliation{Department of Astronomy and Astrophysics, Tata Institute of Fundamental Research,
Mumbai 400005, India}

\author{Achamveedu Gopakumar}
\affiliation{Department of Astronomy and Astrophysics, Tata Institute of Fundamental Research,
Mumbai 400005, India}

\author{Antoine Klein}
\affiliation{GReCO Institut d'Astrophysique de Paris, UMR 7095 du CNRS, Universit\'e Pierre et Marie Curie, 98 bis boulevard Arago, 75014 Paris, France)}

\author{Philippe Jetzer}
\affiliation{Physik-Institut, Universit\"at Z\"urich,
Winterthurerstrasse 190, 8057 Z\"urich}

\begin{abstract}

We provide an elegant way of solving 
analytically the third post-Newtonian (3PN) accurate Kepler 
equation, associated with the 
3PN-accurate generalized quasi-Keplerian parametrization for compact binaries
in eccentric orbits. An additional analytic solution is presented to check the correctness of our compact 
solution and we perform comparisons between our PN-accurate analytic solution and a very accurate numerical solution 
of the PN-accurate Kepler equation.
We adapt our approach to compute crucial 3PN-accurate inputs
that will be required to compute analytically 
both the time and frequency domain  ready-to-use amplitude-corrected 
PN-accurate search templates for compact binaries in inspiralling eccentric orbits.
 
\end{abstract}


\maketitle

\section{Introduction}

The emerging field of gravitational wave (GW) astronomy is expected to mature 
in the coming years and decades.
This expectation is mainly due to the direct detection of GW signals, labeled GW150914 and 
GW151226 \cite{GW_1_d,GW_2_d}, from the coalescence of two distinct binary black hole (BH) systems 
during the first observing run $(O1)$ of the advanced LIGO interferometer \cite{aLIGO}.
The astounding success of LISA pathfinder and maturing pulsar timing arrays ensure that multiwavelength
GW astronomy will be achieved in the coming decades \cite{LPF,PTA}.
Additionally, the coming years are expected to witness a substantial number of 
GW events due to the maturing of a network of ground-based GW
observatories \cite{LRR_LV,O1BBH}.
Coalescing BH binaries in quasicircular orbits should be the dominant
GW sources for these observatories \cite{Rates2016,O1BBH,LRR_LV}.
Preliminary investigations associated with the GW150914 event suggested that
residual eccentricities $\leq 0.1$ at $10$ Hz would not introduce
measurable deviations from the observed GW signal, modeled to be 
from a coalescing BH binary inspiralling along quasicircular orbits \cite{GW_1_prop}.
Indeed, a recent effort shows that 
BH binaries associated with the transient GW events GW150914 and GW151226 are likely to have orbital 
eccentricities below $0.15$ and $0.1$ at the GW frequency of $14$ Hz \cite{eIMR}.
However, there exist a number of astrophysically feasible scenarios
in which binary BH systems can have moderate values of orbital
eccentricities when their GWs enter observatories like aLIGO,
as noted in Refs.~\cite{eIMR,Tiwari_16_eBBH}.

There are ongoing efforts to model GWs associated with eccentric binary BH mergers \cite{Hinder_2010,BDe_2012,eIMR}.
It is customary to employ the phasing prescription, 
developed in Refs.~\cite{DGI,KG06}, for describing the 
inspiral part of eccentric binary coalescence.
This approach extends the early computations 
of Refs.~\cite{PM63,Peters64} by 
incorporating in an efficient manner the effects of 
three timescales that are crucial to 
describe GWs from eccentric inspirals.
The presence of three {\it distinct timescales} are
essentially due to the use 
of the post-Newtonian (PN) approximation to describe the dynamics 
of these binaries. 
In the PN approximation, 
one invokes a certain  gauge-invariant 
dimensionless 
parameter, namely 
$x \equiv { \left( \frac{G m \omega} {c^3} \right) }^{2/3}$,
where $m$ is the total binary mass while $\omega$ stands 
for the orbital (angular) frequency, 
as the expansion parameter.
The use of $x$ is predominant
while expressing 
the frequency and phase evolution of GWs from 
compact binaries as well as the amplitudes of their two polarization
states $h_{+}$ and $h_{\times}$ \cite{LR_LB}.
Let us recall that 
these three distinct timescales are 
associated with that of the orbital motion, 
periastron precession and radiation-reaction effects.
In the GW phasing formalism of Refs.~\cite{DGI,KG06}, one 
models temporal variations in $h_+$ and $h_\times$ that occur
at the orbital and periastron precession timescales in a semianalytical manner.
This is possible due to the availability of a Keplerian-type 
parametric solution to the PN-accurate orbital dynamics of compact binaries.
This solution provides a semianalytical description of 
the precessing eccentric orbits that are associated with 
the PN-accurate dynamics of compact binaries in noncircular orbits
\cite{Memmesheimer2004}.

The present paper provides an elegant analytical solution to the PN-accurate 
Kepler equation associated with the 3PN accurate generalized 
quasi-Keplerian parametrization, available in Ref.~\cite{Memmesheimer2004}.
Specifically, we derive analytical 3PN-accurate infinite series expression for the eccentric 
anomaly $u$ in terms of the mean anomaly $l$.
This solution requires us to derive 
compact PN-accurate infinite series expressions for certain trigonometric functions 
of the true anomaly $v$ in terms of $u$.
We manipulate complex exponential representations of various trigonometric 
functions of $v$ and $u$ for these derivations.
Another analytical solution 
to the 3PN-accurate Kepler equation is also provided 
to check the correctness of our 
solution. We invoke an improved version of
Mikkola's method, detailed in Refs.~\cite{M87,TG06},
to compare the accuracy of our analytical solution for various values 
of the orbital eccentricity.
Our PN-accurate analytic solution shows excellent agreement 
with its numerical counterpart for moderate values of
eccentricity.

We adapt the above computations to derive 3PN-accurate relations between 
various trigonometric functions 
of $v$ and $u$ in terms of $l$.
These relations will be required to compute analytically the time-domain 
response function 
of GW observatories to eccentric inspirals.
One requires 
PN-accurate amplitude-corrected $h_{\times}(t)$ and $h_+(t)$ expressions 
to obtain such ready-to-use response functions, namely
$h(t)= F_{\times} \, h_{\times}(t) + F_+\,h_+ (t)$,
where $F_{\times}$ and $F_+$ are the so-called beam pattern functions of GW 
observatories. It is the practice of expressing 
 $h_{\times}(t)$ and $h_+(t)$  
as sums over various harmonics in $l$, as evident from Eqs.~(3.3)-(3.10) in Ref.~\cite{YABW},
that demands PN-accurate trigonometric functions 
 of $v$ and $u$ in terms of the mean anomaly $l$.
Note that the equations of Ref.~\cite{YABW} provide quadrupolar order GW polarization states associated 
with compact binaries moving along typical Keplerian (or Newtonian) eccentric orbits
and require a solution to the classic Kepler equation and its subsidiary 
results.
Our solution and the associated PN-accurate relations  will be required to extend 
the results of Ref.~\cite{YABW} to 3PN order.
We demonstrate the use of our PN-accurate relations by computing 
analytic 1PN-accurate amplitude-corrected expressions
for $h_{+,\times}(l)$ that are accurate to leading order in orbital eccentricity.

Our prescription to compute analytic amplitude-corrected $h_{+,\times}(l)$ 
will also be required to obtain ready-to-use frequency domain GW response function 
for moderate eccentric inspirals.
This ongoing effort is extending detailed computations, presented in 
Ref.~\cite{THG16},
with the help of the postcircular expansion of PN-accurate eccentric orbits 
and the stationary phase approximation, detailed in Ref.~\cite{YABW}.

In what follows, we sketch the derivation of a popular solution to 
the classic Kepler equation and provide its natural and elegant extension to tackle the 3PN-accurate Kepler 
equation. 
An equivalent but lengthy expression, influenced by Ref.~\cite{Tessmer2011}, 
is presented in Appendix~\ref{sec: alternative solution} while 
Appendix~\ref{sec: trigonometric relations} provides the derivation of 
some of the crucial ingredients that are required for our analytic solution of the 3PN-accurate Kepler equation.
We perform comparisons of our 3PN-accurate analytic 
solution to its numerical counterpart in a subsection of Sec.~\ref{3PN_sol}.
Section~\ref{waveform_Fourier} presents our approach to obtain 
PN-accurate postcircular expansion of time-domain GW polarization states and we discuss its implications.
Many detailed expressions, required for such an effort, and their brief derivations are provided in 
Appendices~\ref{sec: true anomaly},~\ref{sec: Product of Fourier series} and
~\ref{sec: Fourier series of exp(imW)}. Appendix~\ref{sec: h_+,x} provides
1PN amplitude-corrected $h_{+,\times}$ expressions which extend the quadrupolar 
expressions of Ref.~\cite{Wahlquist87}.

\newpage
\section{Derivation of analytic solution to PN-accurate Kepler Equation}
\label{3PN_sol}

We begin by sketching how F.~W.~Bessel invoked his now famous Bessel function to 
solve a demanding transcendental equation proposed by Johannes Kepler \cite{KE_Book}.
An elegant extension of Bessel's approach to solve the 3PN-accurate Kepler equation is presented 
in Sec.~\ref{sec: 3PN-accurate solution} and we probe its numerical accuracy in Sec.~\ref{sec: numerical}.

\subsection{The Bessel function approach to tackle the classic Kepler equation} 

We begin by reviewing the classical  Keplerian parametrization 
that describes semianalytically the Newtonian-accurate orbital motion 
of a binary in noncircular orbits \cite{KE_Book,DD1985}.
In polar coordinates and in the center-of-mass reference frame,
this approach provides a parametric description for an eccentric orbit
of Newtonian dynamics using 
\begin{subequations}
	\begin{align}
		r &= a(1-e\cos u )\,,\\
		\phi -\phi_0 &=v\equiv 2 \arctan \biggl [ \biggl ( \frac{ 1 + e}{ 1 - e}
		\biggr )^{1/2} \, \tan \frac{u}{2} \biggr ]\,,
	\end{align}
\end{subequations}
where $r$ and $\phi$ define the components of the relative separation vector 
${\bf r} = r ( \cos \phi, \sin \phi, 0)$. 
In the above equations, $a$ and $e$ stand for 
the semimajor axis and the eccentricity of the orbit, respectively.
The auxiliary angles $u$ and $v$ are
called eccentric and true anomaly.
The classical Kepler equation defines the temporal evolution of these 
auxiliary angles and is given by
\begin{align}
	l \equiv n (t - t_0)  = u - e\,\sin u\,,
\end{align}
where $l$ is the mean anomaly and
the mean motion $n$ is defined as 
$n = \frac{2\,\pi}{P}$, $P$ being the orbital period.
The quantities $t_0$ and $\phi_0$ are some initial time and 
associated orbital phase.
The conservative nature of the Newtonian orbital 
dynamics allows one to express 
the orbital elements $a$, $e$ and 
$n$ in terms of the Newtonian
orbital energy  and angular momentum.
These expressions are given by
\begin{subequations}
	\begin{align}
		a &= \frac{G\,m}{{(-2\,E)}}\,,\\
		e^2 &= 1 + 2\,E\,h^2\,, \\
		n&= \frac{{{(-2\,E)}}^{3/2}}{Gm}\,, 
	\end{align}
\end{subequations}
where $E$ is the Newtonian orbital energy 
per unit reduced mass $\mu = m_1\,m_2/m$, 
$m_1$ and $m_2$ being the individual masses of the binary and $m= m_1+m_2$.
The scaled angular momentum is given by $h = \frac{J}{G\,m}$, where
$J$ is the reduced Newtonian orbital angular momentum.   
    
Analytic solutions of the classical Kepler equation, namely
$l = u - e\,\sin u$, 
had attracted the attention of 
several generations of distinguished mathematicians 
during the 
nineteenth and twentieth centuries \cite{KE_Book}.
In what follows, we sketch the derivation 
of the widely used solution involving the Bessel functions  
\cite{Watson1922}.

We start by expressing $u-l$ as a Fourier series in $l$:
\begin{align}\label{eq: umlN}
	u - l = e\,\sin u = \sum_{s = 1}^{\infty} A_s \sin(s l) \,,
\end{align}
where the  coefficients $A_s$ are given by 
\begin{align}
	A_s = \frac{2}{\pi} \int_{0}^{\pi} (u-l) \sin(s l) dl\,.
\end{align}
Integrating by parts leads to 
\begin{align}
	A_s =& \frac{2}{\pi} \int_{0}^{\pi} (u(l)-l) \sin(s l) dl \nonumber\\
	=& \frac{2}{s\pi} \int_{0}^{\pi} \cos(s l) du \nonumber\\
	=& \frac{2}{s}
	\biggl \{ \frac{1}{\pi}\,
	\int_{0}^{\pi} \cos(s u - s e \sin u) du  \biggr \}\,.
\end{align}
The expression in the curly brackets can be identified with 
$J_{s}(se)$, namely the Bessel functions of the first kind.
This allows us to write  
\begin{align}\label{eq: N_KE_Solution}
	u = l + \sum_{s=1}^{\infty} \frac{2}{s} J_s(s e) \sin(s l)\,.
\end{align}
This expression provides the most popular solution of the  transcendental Kepler equation.
In what follows, we adapt a similar approach to tackle the PN-accurate Kepler equation.

\subsection{3PN-accurate solution to PN-accurate Kepler equation} \label{sec: 3PN-accurate solution}

The post-Newtonian approach, heavily used to describe dynamics of astrophysical 
systems, incorporates  general relativistic effects as 
perturbations to Newtonian dynamics.
Einstein himself invoked the PN approach for describing 
the perihelion advance of Mercury \cite{Einstein1916}.
We may treat the PN approximation as a computational tool
for tackling the nonlinear
Einsteinian prescription for gravity in terms of certain perturbative deviations 
from the linear Newtonian gravity.
This approach involves  an expansion  
in terms of a small parameter that is usually the squared ratio of the velocity of the matter distribution
forming the gravitational field to the speed of light. For the inspiral dynamics of compact binaries
this small parameter is equivalent to the above defined parameter $x$.
At present, dynamics of compact binaries have been computed to the fourth PN order
which provides general relativity based corrections to Newtonian description
that are accurate to $x^4$ order (see Refs.~\cite{Porto17,DJ17,Foffa16,BBBFM16,DJS2016} and references therein
for the details of this herculean effort from various approaches).

Remarkably, it is possible to obtain a Keplerian-type parametric 
solution to the PN-accurate orbital dynamics of compact binaries
in noncircular orbits \cite{DD1985,DS1988,SW93,Memmesheimer2004}.
At the third post-Newtonian order, the conservative orbital 
dynamics of compact binaries in eccentric orbits is specified
by providing the following parametrization for the   
dynamical variables
$r$ and $\phi$:
\begin{subequations}
\label{eq: PhasingFinalParam3PNharmonic}
	\begin{align}
		r & = a_r \left( 1 - e_r \cos u \right)
		\,,
		\\
		\phi - \phi_{0} & 
		= (1 + k ) v 
		+ \left(f_{4\phi} + f_{6\phi} \right) \sin (2v)
		\nonumber
		\\
		& \quad
		+ \left( g_{4\phi} + g_{6\phi} \right) \sin (3v)
		+  i_{6\phi}\sin (4v)
		\nonumber
		\\
		& \quad
		+ h_{6\phi} \sin (5v) 
		\,,
		\\
		\label{phasing_eq:9c}
		\text{where} \quad
		v & = 2 \arctan 
		\left[ 
		\left( \frac{ 1 + e_{\phi} }{ 1 - e_{\phi} } \right)^{1/2} 
		\tan \frac{u}{2} 
		\right]
		\,.
	\end{align}
\end{subequations}
A distinctive feature of the above two equations is the presence of 
different eccentricity parameters $e_r$ and $e_{\phi}$ for the radial and angular 
variables.
These  were introduced
so that the PN-accurate parametrization looks ``Keplerian" even at higher PN 
orders.
The quantity $k$  provides the rate of periastron advance 
per orbital revolution.
In the above equations, 
$a_r$, $e_r$, and $e_{\phi}$ are
some 3PN accurate semimajor axis,
radial eccentricity, and 
angular eccentricity, while 
$f_{4\phi}$, $f_{6\phi}$, 
$g_{4\phi}$, $g_{6\phi}$, 
$i_{6\phi}$, and 
$h_{6\phi}$
are some orbital functions of the energy and the angular 
momentum that enter at 2PN and 3PN orders.
The explicit PN-accurate expressions of these quantities
are available in Ref.~\cite{Memmesheimer2004}.

The following 3PN accurate Kepler equation
links the eccentric anomaly $u$ 
to the mean anomaly $l = n \left( t - t_0 \right)$
\begin{align}
	\label{eq: 3PN_KE}
	l =&\; u - e_t \sin u + \left(g_{4t} + g_{6t} \right)(v-u) \nonumber\\
		&+ \left(f_{4t} + f_{6t} \right)\sin v + i_{6t} \sin (2v) + h_{6t} \sin (3v)\,.
\end{align}
This PN-accurate Kepler equation requires another 
eccentricity parameter, namely $e_t$,
which is usually called the time eccentricity.
Additionally, there are more orbital functions
$g_{4t}$, $g_{6t}$, 
$f_{4t}$, $f_{6t}$, 
$i_{6t}$, and 
$h_{6t}$ that appear at 2PN and 3PN orders.
The above-mentioned orbital elements 
and functions,
expressible in terms of the conserved orbital energy,
angular momentum, $m$ and $\eta$,  are listed in Ref.~\cite{Memmesheimer2004}.
We observe that the above parametric solution is usually 
referred to as the
``generalized quasi-Keplerian" parametrization 
associated with the 3PN-accurate orbital dynamics.
This is mainly due to the presence of these 
orbital functions that appear at 2PN and 3PN orders.

In what follows, we derive an elegant solution to the
3PN accurate Kepler equation, namely Eq.~(\ref{eq: 3PN_KE}).
It is possible to bring in a compact infinite series expansion, similar to 
Eq.~(\ref{eq: umlN}),
by invoking the following exact relations (see 
Appendix~\ref{sec: trigonometric relations} for 
their  derivations):
\begin{subequations}
\label{eq:3PN_relations}
\begin{align}
	v-u =&\; 2\sum_{j = 1}^{\infty} \frac{\beta_{\phi}^j}{j}\sin(j u) \,,\\
	\sin v =&\; \frac{2 \sqrt{1-e_{\phi}^2}}{e_{\phi}}\sum_{j=1}^{\infty}\beta_{\phi}^j \sin(j u) \,, \\
	\sin (2v) =&\; \frac{4 \sqrt{1-e_{\phi}^2}}{e_{\phi}^2} \sum_{j=1}^{\infty} \beta_{\phi}^j \left(j\sqrt{1-e_{\phi}^2} -1\right) \sin(ju) \,, \\
	\sin (3v) =&\; \frac{2 \sqrt{1-e_{\phi}^2}}{e_{\phi}^3} \sum_{j=1}^{\infty} \beta_{\phi}^j \Big(4-e_{\phi}^2\nonumber\\
	&-6j\sqrt{1-e_{\phi}^2} + 2j^2(1-e_{\phi}^2)\Big) \sin(ju) \,,
\end{align}
\end{subequations}
with $\beta_{\phi} = (1-\sqrt{1-e_{\phi}^2})/e_{\phi}$.
These compact expressions allow us to express Eq.~(\ref{eq: 3PN_KE}) as 
\begin{align}\label{eq: GKE}
	l = u - e_t \sin u + \sum_{j=1}^{\infty} \alpha_j \sin(ju)\,,
\end{align}
where the explicit expressions for the PN-accurate orbital functions $\alpha_j$ 
can be extracted with the help of Eqs.~(\ref{eq: 3PN_KE}) and (\ref{eq:3PN_relations}).
They are given by
\begin{align}\label{eq: alpha3PN}
	\alpha_j =& 2\beta_{\phi}^j \frac{\sqrt{1-e_{\phi}^2}}{e_{\phi}^3} \Bigg((f_{4t} 
	+ f_{6t}) e_{\phi}^2 +\frac{(g_{4t} + g_{6t}) e_{\phi}^3}{j\sqrt{1-e_{\phi}^2}} \nonumber\\
	&+ 2i_{6t}e_{\phi} \left[j\sqrt{1-e_{\phi}^2}-1\right] \nonumber\\
	&+ h_{6t} \left[4-e_{\phi}^2-6j\sqrt{1-e_{\phi}^2} + 2j^2(1-e_{\phi}^2)\right] \Bigg)\,.
\end{align}
It is worth noting that the 
functional forms of $\alpha_j$ are identical in both the modified 
harmonic (MH) and Arnowitt-Deser-Misner (ADM) coordinates, since
Eq.~\eqref{eq: 3PN_KE} takes an identical form in both 
gauges~\cite{Memmesheimer2004}.
However, the explicit expressions for these orbital functions in terms of the 
conserved orbital energy and angular momentum or the parameters
$x$ and $e_t$ differ.

The functional form of the PN-accurate Kepler equation, namely 
Eq.~(\ref{eq: GKE}), allows us to write 
the following PN-accurate Fourier series for $u-l$ 
\begin{align}
	u - l = \sum_{s = 1}^{\infty} A_s \sin(s l) \,,
\end{align}
where the  coefficients $A_s$ are defined as
\begin{align}
	A_s = \frac{2}{\pi} \int_{0}^{\pi} (u-l) \sin(s l) dl\,.
\end{align}
Integrating by parts and using Eq.~(\ref{eq: GKE}) gives
\begin{align}
	A_s =&\; \frac{2}{\pi} \int_{0}^{\pi} (u-l) \sin(s l) dl \nonumber\\
	=&\; \frac{2}{s\pi} \int_{0}^{\pi} \cos(s l) du \nonumber\\
	=&\; \frac{2}{s\pi} \int_{0}^{\pi} \cos\Big(s u - s e_t \sin u 
	+ s \sum_{j=1}^{\infty} \alpha_j \sin(ju) \Big) du \,.
\end{align}
Note that the $\alpha_j$ contributions appear 
only at 2PN and 3PN orders as evident from Eq.~(\ref{eq: alpha3PN}).
Therefore, we expand the sum in the cosine function of the above 
integral to the first order in $\alpha_j$. This leads to
\begin{align}
	A_s =&\; \frac{2}{s\pi} \int_{0}^{\pi} \cos\left(s u - s e_t \sin u \right) du \nonumber\\
	&- \frac{2}{\pi} \sum_{j=1}^{\infty} \alpha_j \int_{0}^{\pi} \sin\left(s u - s e_t \sin u \right)\sin(ju) du \nonumber\\
	=&\; \frac{2}{s\pi} \int_{0}^{\pi} \cos\left(s u - s e_t \sin u \right) du \nonumber\\
	&+ \frac{1}{\pi} \sum_{j=1}^{\infty} \alpha_j \int_{0}^{\pi} \big\{\cos\left((s+j) u - s e_t \sin u\right) \nonumber\\
	&- \cos\left((s-j) u - s e_t \sin u\right) \big\} du \nonumber\\
	=&\; \frac{2}{s}  J_s(s e_t) + \sum_{j=1}^{\infty} \alpha_j \left\{ J_{s+j}(s e_t) - J_{s-j}(s e_t) \right\} \,,
\end{align}
where we employed the usual integral definitions for $J_n(x)$
to reach the last step.
This step allows us to write down a simple and elegant solution to 3PN-accurate generalized Kepler equation
in terms of Bessel functions as
\begin{subequations}\label{eq: KE solution}
	\begin{align}
		u &= l + \sum_{s=1}^{\infty} A_s \sin(sl) \,,\\
		A_s &= \frac{2}{s}  J_s(s e_t) + \sum_{j=1}^{\infty} \alpha_j \left\{ J_{s+j}(s e_t) - J_{s-j}(s e_t) \right\} \,.
	\end{align}
\end{subequations}
Clearly, one requires explicit expressions for $\alpha_j$
in terms of $x$, $e_t$ and $\eta$ while employing our solution. 
The relevant expressions, valid for MH and ADM gauges,
may be computed from Ref.~\cite{Memmesheimer2004} as 
\begin{widetext}
	\begin{subequations}
		\begin{align}
			\alpha_j^{\rm H} =&\; \beta_t^j \Bigg\{ x^2 \Bigg( \frac{15-6\eta}{j \sqrt{1-e_t^2}} + \frac{15\eta - \eta^2}{4}\Bigg) 
			+ x^3 \Bigg( \frac{2880\left(1+e_t^2\right)-\left(10880 + 2784 e_t^2 -123\pi^2\right)\eta + \left(960 + 1056e_t^2\right)\eta^2}{96 j \left(1-e_t^2\right)^{3/2}} \nonumber \\
			&+ \frac{268800 - \left(182192 + 1120 e_t^2 + 4305 \pi^2 \right)\eta + \left(8260 - 11620 e_t^2\right)\eta^2 - 1820\left(1-e_t^2\right) \eta^3}{3360\left(1-e_t^2\right)} \nonumber \\
			&+ j \frac{681 \eta - 199\eta^2 + 3 \eta^3}{16 \sqrt{1-e_t^2}} + j^2 \frac{23\eta-73\eta^2 + 13\eta^3}{48} \Bigg)	\Bigg\} \,,\\
			\alpha_j^{\rm A}=&\; \beta_t^j \Bigg\{ x^2 \Bigg( \frac{15-6\eta}{j \sqrt{1-e_t^2}} - \frac{4\eta + \eta^2}{4}\Bigg)
			+ x^3 \Bigg( \frac{2880\left(1+e_t^2\right)-\left(10880 + 2784 e_t^2 - 123\pi^2\right)\eta + \left(960 + 1056e_t^2\right)\eta^2}{96 j \left(1-e_t^2\right)^{3/2}} \nonumber \\
			&+ \frac{7488 - \left(7544 - 48 e_t^2 - 3 \pi^2 \right)\eta + \left(1168 + 32 e_t^2\right)\eta^2 - 52\left(1-e_t^2\right) \eta^3}{96\left(1-e_t^2\right)} \nonumber \\
			&+ j \frac{-18 \eta + 24\eta^2 + 3 \eta^3}{16 \sqrt{1-e_t^2}} + j^2 \frac{13\eta^3}{48} \Bigg)	\Bigg\} \,,
		\end{align}
	\end{subequations}
\end{widetext}
where the superscripts ${\rm H}$ and ${\rm A}$  stand for the two gauges 
involved, namely the MH and ADM gauges. 
We note that $\beta_t = (1-\sqrt{1-e_t^2})/e_t$ is defined with the time eccentricity.
To provide a check on our PN-accurate solution, we derive 
in Appendix~\ref{sec: alternative solution}  an alternate 
and less compact solution to the 3PN-accurate Kepler equation that is 
influenced by Ref.~\cite{Tessmer2011}.
We expand our two 3PN-accurate solutions to ${\cal O}(e_t^{40})$ to verify that they are identical
at each order in $e_t$.

In what follows, we compare our solution with the 
2PN-accurate $u(l)$ solution of Ref.~\cite{Tessmer2011}.
This solution in our notation reads 
\begin{subequations}
	\label{eq: KE Tessmer2011}
	\begin{align}
		u =&\; l + \sum_{s=1}^{\infty} A_s \sin(sl) \,,\\
		A_s =&\; \frac{2}{s}J_s(s e_t)\nonumber\\
		&-\sum_{j=1}^{\infty}\alpha_j\left\{J_{j-s}((j-s) e_t) - J_{j+s}((j+s)e_t) \right\} \,,
	\end{align}
\end{subequations}
with the constant coefficients $\alpha_j$ given by
\begin{align}\label{eq: alphaj Tessmer2011}
	\alpha_j =&\; \frac{2\;g_{4t}}{j} \sum_{i=1}^{\infty} \beta_{\phi}^i\left\{J_{j-i}(j e_t) + J_{j+i}(j e_t) \right\} \nonumber\\
	& + f_{4t} \sqrt{1-e_t^2}\left\{J_{j-1}(j e_t) - J_{j+1}(j e_t) \right\} \,.
\end{align}
We observe that two typos are persistent in Ref.~\cite{Tessmer2011} while trying to express 
$\sin v$ in terms of $l$.
This is evident by comparing their Eq.~(87) with our Eq.~(\ref{eq: Fourier v}) or its equivalent
that may be found in a classical treatise like Ref.~\cite{Watson1922}.
Additionally, the arguments of the Bessel functions should read
$(k-n)e_t$ and $(k+n)e_t$ while going from steps 7 to 8 of Eq.~(149) in Ref.~\cite{Tessmer2011}. 
These corrections ensure that Eq.~(\ref{eq: KE Tessmer2011}) is consistent with our elegant solution
at 2PN order.
To check the consistency of these two solutions, 
we expand Eqs.~(\ref{eq: KE solution}) and (\ref{eq: KE Tessmer2011}) around $e_t=0$.
We have verified that they are in perfect agreement up to 
$\mathcal{O}(e_t^{40})$.

We observe that the approach of Ref.~\cite{Tessmer2011} results in a complicated PN-accurate expression for
$u(l)$ as evident from our Eqs.~(\ref{eq: KE Tessmer2011}) and (\ref{eq: alphaj Tessmer2011}).
This is mainly due to the presence of infinite Bessel series in the constant $\alpha_j$.
It turned out to be rather difficult to extend the prescription of Ref.~\cite{Tessmer2011}
to 3PN order. This prompted us 
to develop a 3PN extension of Eq.~(\ref{eq: KE Tessmer2011}) that 
requires PN-accurate compact relations, given by our Eqs.~(\ref{eq:3PN_relations}).
This additional solution,
detailed in Appendix~\ref{sec: alternative solution}, provided an independent check for 
our 3PN-accurate elegant solution. 

\subsection{Comparison to numerical solution}\label{sec: numerical}

\begin{centering}
	\begin{figure}[!h]
		\includegraphics[trim={0 1cm 0 2cm},clip,scale=0.3]{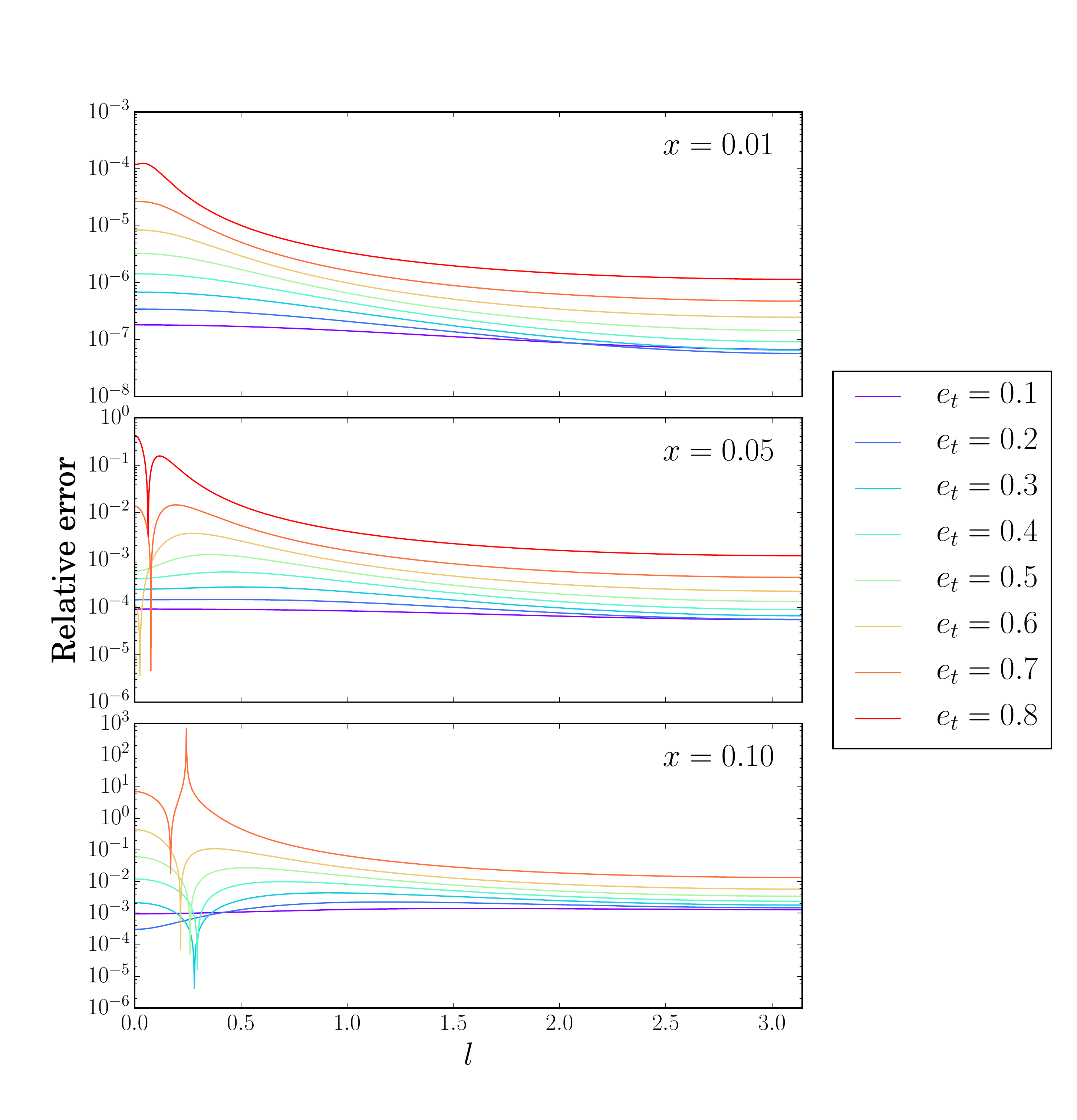}
		\vspace{-0.75cm}
		\caption{
		The fractional relative error $|((u_{\textnormal{num}}-l)-(u_{\textnormal{anl}}-l))/(u_{\textnormal{num}}-l)|$ as a function of the mean anomaly $l$ for different $e_t$ and $x$ values. We let $\eta = 0.25$ and truncate the analytic series solution at $j=200$. $x \approx 0.01$ corresponds to a binary neutron star system entering the aLIGO band at 10 Hz, while a binary black hole system with masses around $10 M_\odot$ enters at $x \approx 0.03$.}
		\label{fig: relative_error_x}
	\end{figure}
\end{centering}

In this subsection we compare our analytic solution against a very accurate way of solving the PN-accurate 
Kepler equation, detailed in Refs ~\cite{TG07,THG16}.
This numerical approach is based on an efficient and accurate (numerical) way of solving the
classical Kepler equation, developed by Mikkola \cite{M87} and 
is valid for all $l$ and for $0 \leq e_t \leq 1$.
Mikkola's method involves finding an analytic solution to 
certain cubic polynomial and a subsequent 
fourth-order iteration to improve on the initial guess for $u$.
Its PN extension involves iteratively invoking the method 
to tackle PN-accurate Kepler equation, expressed in certain 
``quasiclassical" form (see Refs.~\cite{TG06,THG16} for details).
We observe that the PN-accurate analytic solution is fully specified by providing 
values for $l$, $e_t$, $x$, and $\eta$.
Our analytic solution is expected to be valid only up to certain values 
of the PN-expansion parameter $x$ and it will diverge for large values of $x$.
Additionally, it will be useful to concentrate on the 
differences between $u$ and $l$ values  due to the nature of Eq.~(\ref{eq: KE solution}).
These considerations influenced us to probe how the fractional relative error,
namely $|((u_{\textnormal{num}}-l)-(u_{\textnormal{anl}}-l))/(u_{\textnormal{num}}-l)|$,
varies as a function of $e_t$ for few $x$ values while 
incorporating $200$ terms in the analytic solution.
The results in MH gauge, displayed in Fig.~\ref{fig: relative_error_x}, 
reveal that the relative error is small 
for  moderate eccentricities and reasonable $x$ values.
However, this error estimate can approach unity for $x$ values like $0.1$ 
even with moderate eccentricities ($e_t = 0.7$).
In any case, the maximum factional relative error is below
$10\%$ for $e_t < 0.5$ and $x=0.1$ for equal mass compact binaries.
  
\begin{centering}
	\begin{figure}[!h]
		\includegraphics[trim={3cm 0.5cm 0 2cm},clip,scale=0.16]{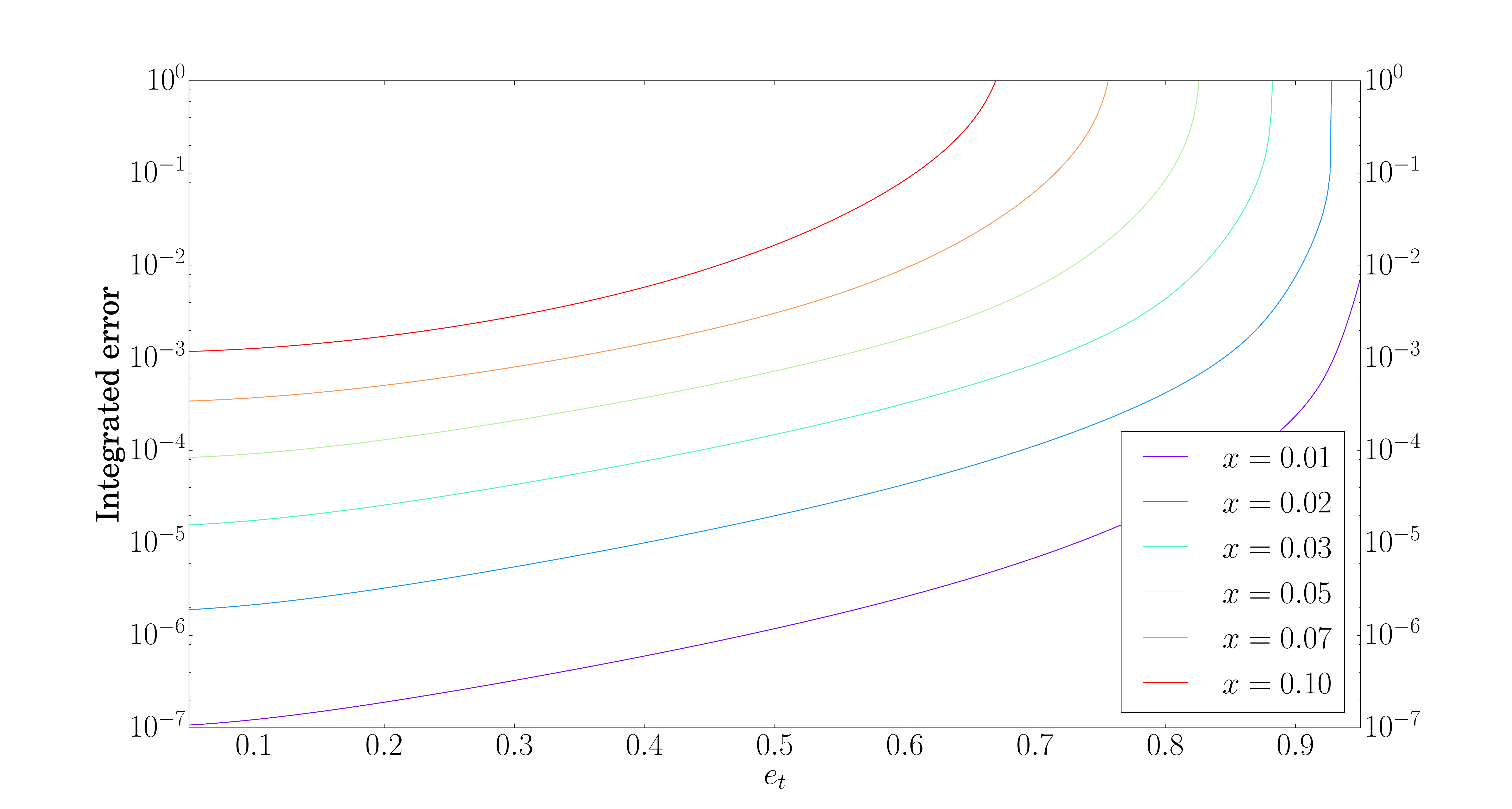}
		\vspace{-0.75cm}
		\caption{Integrated relative error as a function of $e_t$ for different $x$ values.
		The other parameters are as in Fig.~\ref{fig: relative_error_x}.}
		\label{fig: integrated_rel_error_x}
	\end{figure}
\end{centering}

We invoke the more familiar integrated error over one period using the $L^2$ norm, namely
\begin{align}
	\lVert f \rVert_{L^2} &= \left(\frac{1}{2\pi} \int_{0}^{2\pi} f^2 dl\right)^{1/2} \,,
\end{align}
where $f$ stands for the above-mentioned fractional relative error.
In Fig.~\ref{fig: integrated_rel_error_x}, we show this error estimate as a function of $e_t$ 
for a number of $x$ values.
We find that our  $L^2$ norm error estimate is small ($<1\%$) for eccentricities up to $e_t = 0.95$ for 
$x$ values relevant for the early inspiral phase like $x\approx0.01$.
However, it diverges quickly for higher $x$ values and this is true even for moderate $e_t$ values like $0.5$.
A possible explanation is that 
this behavior happens when $e_{\phi} \sim e_t(1+x(4-\eta)) +\mathcal{O}(x^2)$ approaches unity.
It is easy to infer that this happens when $e_t \approx 1/(1+4x)$ and this is consistent 
with our plots.

In what follows, we introduce a new parameter to specify cleanly 
where our analytic solution is accurate, trustable and devoid of the above divergences.
This post-Newtonian parameter is defined to be 
\begin{align}
	y = \frac{(Gm\omega)^{1/3}}{\sqrt{1-e_t^2}} \,.
\end{align}
It smoothly goes to the standard post-Newtonian parameter $x^{1/2}$ in the circular limit.

\begin{centering}
	\begin{figure}[!h]
		\includegraphics[trim={0 1cm 0 2cm},clip,scale=0.3]{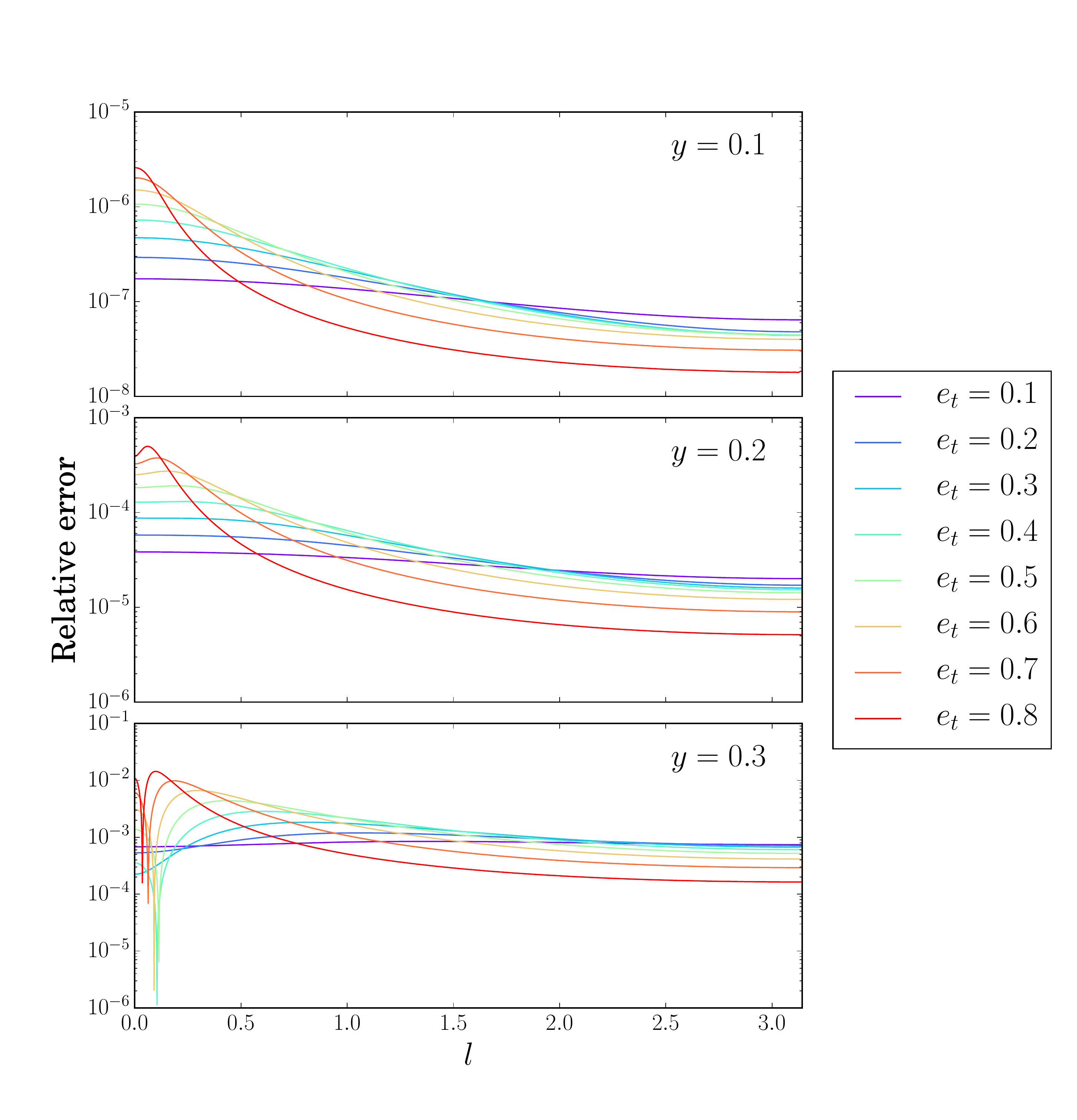}
		\vspace{-0.75cm}
		\caption{Relative error $|((u_{\textnormal{num}}-l)-(u_{\textnormal{anl}}-l))/(u_{\textnormal{num}}-l)|$ as a function of the mean anomaly $l$ for different $e_t$ and $y$ values.
			In the circular limit $y=0.1$ corresponds to $x=0.01$ and $y=0.316$ to $x=0.1$.}
		\label{fig: relative_error_y}
	\end{figure}
\end{centering}

\begin{centering}
	\begin{figure}[!h]
		\includegraphics[trim={3cm 0.5cm 0 2cm},clip,scale=0.16]{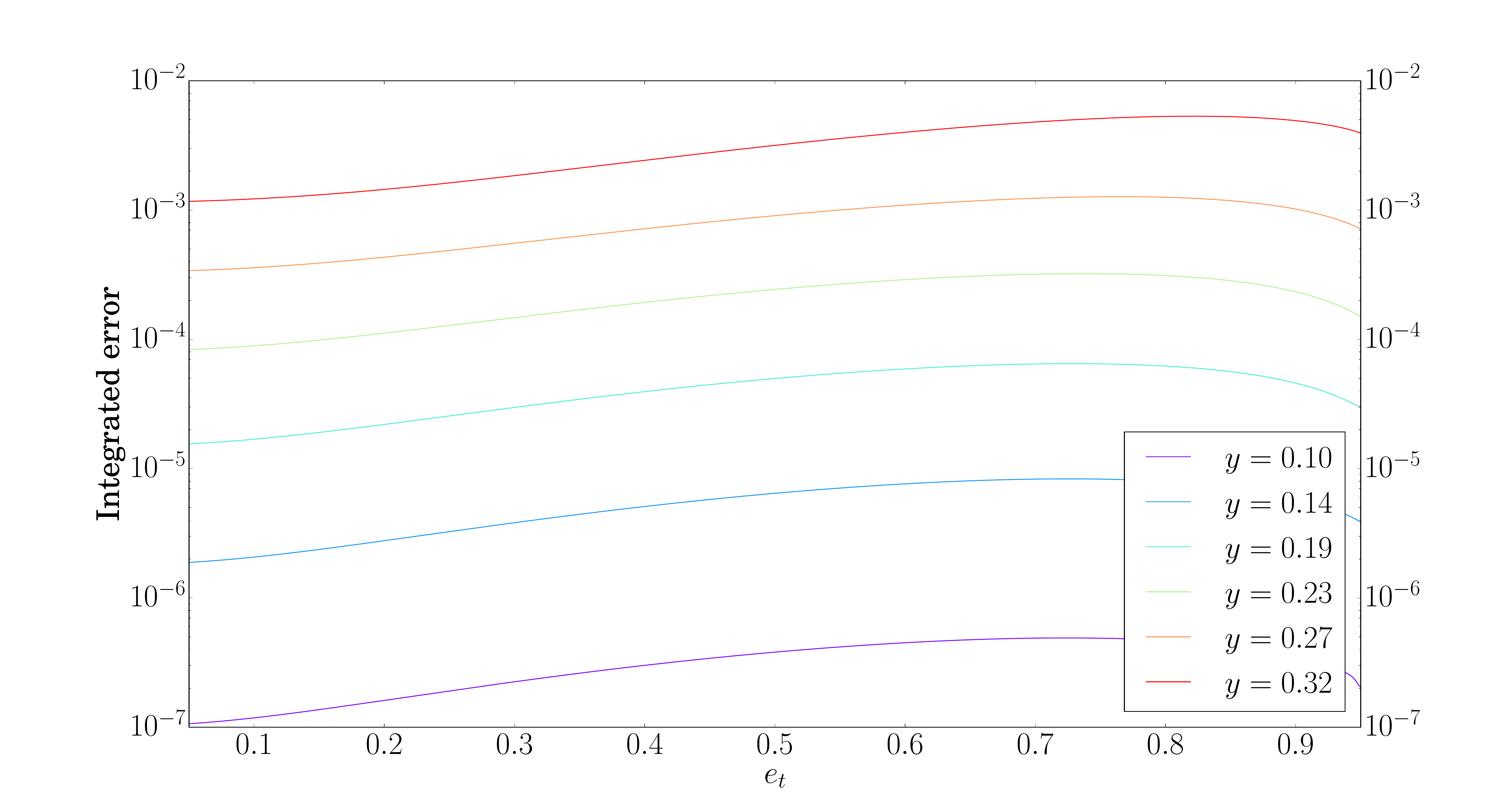}
		\vspace{-0.75cm}
		\caption{Integrated relative error as a function of the eccentricity $e_t$ for different $y$.}
		\label{fig: integrated_rel_error_y}
	\end{figure}
\end{centering}

We plot in Fig.~\ref{fig: relative_error_y} the 
fractional relative error as a function of $l$ for several $e_t$ and few $y$ values.
The sharp maxima, visible in Fig.~\ref{fig: relative_error_x},
are absent in such $y$ plots and the maximum relative error is less than $1\%$
for large $y$ values like $0.3$.
This is repeated in Fig.~\ref{fig: integrated_rel_error_y} for the integrated error as function of $e_t$ 
for several $y$ values. We again find smooth behavior and noticeably lower error estimates (less than $1\%$)
for high $y$ and $e_t$ values.

In Figs. \ref{fig: relative_error_x} to \ref{fig: integrated_rel_error_y} we only considered equal mass binaries.
We found similar behavior for Neutron star-black hole binaries ($\eta \sim 0.1$).
These estimates suggest that our analytic solution should be accurate to compute analytic PN-accurate 
$h_{+,\times}(l)$ expressions for moderately eccentric inspirals.
This is what we pursue in the next section.

\section{Inputs to compute analytic time-domain amplitude-corrected $h_{+,\times}(l)$}
\label{waveform_Fourier}
In this section we derive inputs that will be required to compute 3PN-accurate amplitude-corrected
expressions for the time-domain $h_{+,\times}$ as a sum over {\it harmonics} in $l$.
These PN-accurate results, as expected, will also be required to obtain amplitude 
corrected Fourier-domain inspiral templates with the help of Refs.~\cite{YABW,THG16}.
Such PN-accurate input expressions can be regarded as nontrivial corollaries to
our analytical solution to the 3PN-accurate Kepler equation.
The various Fourier series coefficients derived in this section
are given in a Mathematica Notebook in the Supplemental Material~\cite{supplement}.

We begin by listing quadrupolar, Newtonian order expressions for $h_{+,\times}$ associated 
with nonspinning compact binaries in eccentric orbits, adapted from \cite{Gopakumar2002,DGI},
\begin{widetext}
	\begin{subequations}\label{eq: hpx_qu}
		\begin{align}	
		h_{+}^{0} =&\, \frac{G\,m\, \eta}{2\,R'\, c^2}\frac{x}{(1-e_t \cos u)^2} \Big\{ s_i^2 \left(-e_t^2 + 2 e_t \cos u -e_t^2 \cos(2u)\right) \nonumber\\ 
					&-\left(1+c_i^2\right) \Big\{ \left[ 4-3e_t^2-2 e_t \cos u + e_t^2 \cos(2u) \right] \cos(2\Phi) - 4 \sin u\, e_t \left(1-e_t^2\right)^{1/2} \sin(2\Phi) \Big\} \Big\} \,,\\
		h_{\times}^{0} =&\, \frac{G\,m\, \eta}{R'\, c^2}\frac{x\, c_i}{(1-e_t \cos u)^2} \left\{ \left[ 4-3e_t^2 -2e_t \cos u + e_t^2 \cos(2u)\right] \sin(2\Phi) 
						 + 4 \sin u\, e_t \left(1-e_t^2\right)^{1/2} \cos(2\Phi) \right\} \,,
		\end{align}
	\end{subequations}
\end{widetext}
where $R'$ is the luminosity distance and $\Phi= \beta - \phi$.
The source direction is specified by
$(\iota, \beta)$ while $c_i= \cos \iota$, $s_i= \sin \iota$.
We introduce $\Phi$ that combines the orbital phase $\phi$ with $\beta$.  
The orbital phase is specified by employing 3PN-accurate 
generalized quasi-Keplerian parametrization and it reads
\begin{align}
	\phi - \phi_0 =&\; (1+k)v + (f_{4\phi}+f_{6\phi}) \sin (2v) \nonumber\\
	&+ (g_{4\phi}+g_{6\phi}) \sin(3v) + i_{6\phi}\sin(4v) + h_{6\phi} \sin(5v) \,.
\end{align}
It is customary to split $\phi$ into an angle $\lambda$, which is linear in $l$, and 
$W(l)$, which is $2\pi$ periodic in $l$ \cite{Gopakumar2002,DGI}. This allows us to write 
\begin{subequations}
	\begin{align}
		\phi  =&\; \lambda + W(l) \,,\\
		\lambda =&\; \phi_0 + (1+k)l \,,\\
		W(l) =&\; (1+k)(v-l) + (f_{4\phi}+f_{6\phi}) \sin(2v) \nonumber\\
		&+ (g_{4\phi}+g_{6\phi}) \sin(3v) + i_{6\phi}\sin(4v) + h_{6\phi} \sin(5v) \,.
	\end{align}
\end{subequations}
This split of $\phi$ is done to incorporate the advance of periastron explicitly 
into the GW phase evolution and its implications are discussed in Refs.~\cite{Gopakumar2002,TG06}.
A close inspection of Eqs.~(\ref{eq: hpx_qu}) reveals that we need to express 
the cosine and sine of $W(l)$ and $(1 -e_t \cos u)^{-2}$ as functions of the mean anomaly
$l$ to obtain $h_{+,\times}$ as a sum over {\it harmonics} in $l$.
It is not very difficult to infer that
the derivations of such series expressions demand additional PN-accurate Fourier series of $\sin(ju)$, $\cos(ju)$, $\sin(jv)$
and $\cos(jv)$. In what follows, we tackle these challenges.

\subsection{PN-accurate Fourier series expressions for various trigonometric functions of $u$,$v$ and $W$}
 
We begin by deriving explicit expressions for the coefficients $\sigma_s^{ju}$ and $\zeta_s^{ju}$ such that
3PN-accurate Fourier series for $\sin (ju)$ and $\cos (ju)$ can be expressed as  
\begin{subequations}
	\begin{align}
		\sin(ju) &= \sum_{s=1}^{\infty} \sigma_s^{ju} \sin(sl) \,,\\
		\cos(ju) &= \sum_{s=0}^{\infty} \zeta_s^{ju} \cos(sl) \,.
	\end{align}
\end{subequations}
We adopt certain $3$ indices notation to keep track of a number of coefficients 
that will be derived in this subsection.
Let us emphasize that both $\sigma_s^{ju}$ and $\zeta_s^{ju}$ are not functions of $u$.
We briefly describe how these Fourier coefficients are calculated in the Keplerian  parametrization.
The Fourier coefficients $\sigma_s^{ju}$ are defined as 
\begin{align}
	\sigma_s^{ju} =&\; \frac{2}{\pi}\int_{0}^{\pi}\sin(ju)\sin(sl) dl \nonumber\\
	=&\; \frac{2j}{\pi s} \int_{0}^{\pi} \cos(ju)\cos(sl)du \nonumber\\
	=&\;\frac{j}{\pi s} \int_{0}^{\pi} \big\{ \cos((s-j)u - s e\sin u) \nonumber\\
	 &+ \cos((s-j)u - s e\sin u) \big\}du \nonumber\\
	 =&\; \frac{j}{s}\left\{ J_{s+j}(s e) + J_{s-j}(s e) \right\} \,,
\end{align}
where we employed the Newtonian Kepler equation $l = u-e\sin u$ and 
invoked the standard integral definition of Bessel functions of the first kind.

To extend it to 3PN order, we write our 
PN-accurate Kepler equation as 
$l = u-e_t\,\sin u + \sum_{j} \alpha_j \sin(ju)$, due to Eq.~(\ref{eq: GKE}).
We adapt the calculation to obtain $A_s$, detailed in  Sec.~\ref{sec: 3PN-accurate solution},
by expanding 
$\cos((s+j)u -se_t\sin u + s\sum_{j} \alpha_j\sin(ju))$ in terms of the small parameters $\alpha_j$. The
resulting 3PN-accurate Fourier series for $\sin(ju)$ reads 
\begin{subequations}
	\begin{align}\label{eq: sinju}
		\sin(ju) =&\; \sum_{s=1}^{\infty}\sigma_s^{ju} \sin(sl) \,,\\
		\sigma_s^{ju} =&\; \frac{j}{s}\left\{ J_{s+j}(s e_t) + J_{s-j}(s e_t)  \right\} \nonumber\\
		&+\frac{j}{2} \sum_{i=1}^{\infty}\alpha_i \big\{ J_{s+j+i}(s e_t) - J_{s+j-i}(s e_t) \nonumber\\
		&+ J_{s-j+i}(s e_t) -J_{s-j-i}(s e_t)\big\}\,.
	\end{align}
\end{subequations}
Following similar steps, we can easily obtain 3PN-accurate Fourier series for $\cos(ju)$ as
\begin{subequations}
	\begin{align}
	\cos(ju) =&\; \sum_{s=0}^{\infty}\zeta_s^{ju} \cos(sl) \,, \\
	\zeta_0^{ju} =&\; \frac{1}{2}\left(-e_t\,\delta_{j1} + \alpha_j j\right) \,,\\
	\zeta_s^{ju} =&\; \frac{j}{s}\left\{ J_{s-j}(s e_t) - J_{s+j}(s e_t)  \right\} \nonumber\\
	&+\frac{j}{2} \sum_{i=1}^{\infty}\alpha_i \big\{ J_{s-j+i}(s e_t) - J_{s-j-i}(s e_t) \nonumber\\
	&- J_{s+j+i}(s e_t) + J_{s+j-i}(s e_t)\big\}\,,
	\end{align}
\end{subequations}
where $\delta_{j1}$ stands for the standard Kronecker delta.
It is possible to provide a compact expression for $e^{iju} $ by combining the above results for 
$\cos(ju)$ and $\sin(ju)$ as  $e^{iju} = \cos(ju) + i\sin(ju)$.
The resulting expression is given by
\begin{subequations}
	\begin{align}
		e^{iju} =& \; \sum_{s=-\infty}^{\infty} \epsilon_s^{ju}\, e^{isl} \,,\\
		\epsilon_0^{ju} =& \; \frac{1}{2}\left(-e_t\,\delta_{j1} + \alpha_j j\right) \,,\\
		\epsilon_s^{ju} =& \; \frac{j}{s} J_{s-j}(s e_t) \nonumber\\
						&+ \frac{j}{2} \sum_{k=1}^{\infty}\alpha_k \left\{J_{s-j+k}(s e_t)-J_{s-j-k}(s e_t)\right\} \,.
	\end{align}
\end{subequations}

We now move to derive the Fourier series of $\sin(jv)$
and $\cos(jv)$ in terms of the mean anomaly $l$ with the
help of the above expressions. The plan is to write down a series expansion 
for $\sin(jv)$ in terms of $u$ as 
\begin{align}
	\sin(jv) &= \sum_{s=1}^{\infty} \mathcal{E}_s^j \sin(s u) \,.
\end{align}
The above form is justified by our computations as detailed in 
Appendix~\ref{sec: trigonometric relations}.
We invoke the Fourier series of $ \sin(ju)$, given by Eq.~(\ref{eq: sinju}), to obtain
\begin{subequations}
	\begin{align}
	\sin(jv) &= \sum_{s=1}^{\infty} \sigma_s^{jv} \sin(sl) \,,\\
	\sigma_s^{jv} &=  \sum_{k=1}^{\infty} \mathcal{E}_k^j \sigma_s^{ku}\,,
	\end{align}
\end{subequations}
where $\mathcal{E}_k^j$ is given by Eqs.~(\ref{eq: e^ijv}).
Following similar arguments, we obtain PN-accurate results for 
 $\cos(jv)$ as 
\begin{subequations}
	\begin{align}
	\cos(jv) &= \sum_{s=0}^{\infty} \zeta_s^{jv} \cos(sl) \,,\\
	\zeta_s^{jv} &=  \sum_{k=0}^{\infty} \mathcal{E}_k^j \zeta_s^{ku}\,,
	\end{align}
\end{subequations}
and $e^{ijv}$ as 
\begin{subequations}\label{eq: Fourier series of e^ijv}
	\begin{align}
	e^{ijv} &= \sum_{s=-\infty}^{\infty}\epsilon_s^{jv} e^{isl} \,,\\
	\epsilon_s^{jv} &= \sum_{k=0}^{\infty} \mathcal{E}_k^j \epsilon_s^{ku}\,.
	\end{align}
\end{subequations}

We are now in a position to derive 3PN-accurate Fourier series expressions  for 
$\cos(mW)$ and $\sin(mW)$.
The starting point of our derivation is the following equation
\begin{subequations}\label{eq: Fourier series of W}
	\begin{align}
		W(l) =&\; \sum_{s=1}^{\infty}\mathcal{W}_s \sin(sl)\,,\\
		\mathcal{W}_s =&\; (1+k) B_s + (f_{4\phi}+f_{6\phi}) \sigma_s^{2v} \nonumber\\
		&+ (g_{4\phi}+g_{6\phi}) \sigma_s^{3v} + i_{6\phi} \sigma_s^{4v} + h_{6\phi} \sigma_s^{5v} \,.
	\end{align}
\end{subequations}
This equation arises from the 3PN-accurate expression for $W(l)$ given in Ref.~\cite{DGI}
\begin{align}
	W(l) =&\; (1+k)(v-l) + (f_{4\phi}+f_{6\phi}) \sin(2v) \nonumber\\
	&+ (g_{4\phi}+g_{6\phi}) \sin(3v) + i_{6\phi}\sin(4v) + h_{6\phi} \sin(5v) \,,
\end{align}
and our earlier derived series expressions for $\sin(jv)$,
as well as a series expression for the true anomaly $v-l$,
derived in Appendix~\ref{sec: true anomaly}.
We list these relevant expressions again 
\begin{subequations}
	\begin{align}
		v - l&= \sum_{s=1}^{\infty} B_s \sin(sl) \,,\\
		\sin(jv) &= \sum_{s=1}^{\infty} \sigma_s^{jv} \sin(sl) \,.
	\end{align}
\end{subequations}
A straightforward computation that employs the above three infinite series 
expressions leads to the following Fourier series of $e^{imW}$  in terms of $l$:
\begin{align} \label{eq: Fourier coefficients e^imW - 1}
	e^{imW} &= \sum_{n=-\infty}^{\infty} \mathcal{P}_n^{mW} e^{inl} \,.
\end{align}
The Fourier coefficients $\mathcal{P}_n^{mW}$ are given in Appendix~\ref{sec: Fourier series of exp(imW)}, where we describe the derivation  
of Eq.~(\ref{eq: Fourier coefficients e^imW - 1}) in detail.
It is then fairly routine to extract Fourier series of $\cos(mW)$  as 
\begin{subequations}
	\begin{align}
		\cos(mW) &= \sum_{n=0}^{\infty}\mathcal{C}_n^{mW}\cos(nl) \,,\\
		\mathcal{C}_0^{mW} &= \mathcal{P}_0^{mW} \,,\\
		\mathcal{C}_n^{mW} &= \mathcal{P}_n^{mW} + \mathcal{P}_{-n}^{mW} \,,
	\end{align}
\end{subequations}
and $\sin(mW)$ is given by
\begin{subequations}
	\begin{align}
	\sin(mW) &= \sum_{n=1}^{\infty}\mathcal{S}_n^{mW}\sin(nl) \,,\\
	\mathcal{S}_n^{mW} &= \mathcal{P}_n^{mW} - \mathcal{P}_{-n}^{mW} \,.
	\end{align}
\end{subequations}

Finally, we turn our attention to the derivation of $(1-e_t\cos u)^{-n}$. We adapt and extend the 
approach of Ref.~\cite{Tessmer2011} to obtain 3PN-accurate Fourier series of $(1-e_t\cos u)^{-n}$.
Adapting the relevant result in Ref.~\cite{Tessmer2011}, we write
\begin{subequations}
	\begin{align}
		\frac{1}{(1-e_t\cos u)^n} =&\; \sum_{j=0}^{\infty}b^n_j \cos(j u)\,,\\
		b^n_0 =&\; {}_2F_1\left(\frac{n}{2}, \frac{n+1}{2}; 1; e_t^2\right) \,,\\
		b^n_j =&\; \frac{e_t^j}{2^{j-1}} \binom{n+j-1}{j} \nonumber\\
		&\times{}_2F_1\left( \frac{n+j}{2}, \frac{n+j+1}{2};j+1;e_t^2 \right) \,,
	\end{align}
\end{subequations}
\vspace{0.1cm} 
where ${}_2F_1$ stands for the ordinary hypergeometric function.
Combining the above expression with the results for $\cos(ju)$, we get a
3PN-accurate Fourier series
for $1/(1-e_t\cos u)^n$ as 
\begin{subequations}
	\begin{align}
		\frac{1}{(1-e_t\cos u)^n} =&\; \sum_{j=0}^{\infty} \mathcal{A}_j^n \cos(j l)\,,\\
		\mathcal{A}_j^n =&\; \sum_{k=0}^{\infty} b_k^n \zeta_j^{ku} \,.
	\end{align}
\end{subequations}

In the next subsection, we apply the 1PN version of these results
to demonstrate their utility in computing 
analytic $h_{+,\times}$ as a sum over {\it harmonics} in $l$.

\vspace{0.5cm} 
\subsection{Analytic $h_{+,\times}(l)$ via small eccentricity expansion}

The plan is to apply the above derived PN-accurate series expansions to compute analytic 
1PN-accurate amplitude-corrected expressions for $h_{+,\times}(l)$ in the small $e_t$ approximation.
We begin from the exact 1PN-accurate amplitude-corrected 
 $h_{+,\times} $ expressions that we symbolically write as  
\begin{align}
	h_{+,\times} &= \frac{G\,m\, \eta}{R'\, c^2}x \left \{ H^0_{+,\times} + x^{0.5}\, H^{0.5}_{+,\times} +  x\, H^1_{+,\times} \right \} \,.
\end{align}
$H^i_{+,\times}$ are functions of $\Phi = \beta - \phi = \beta - ( \lambda + W) $ and $u$.
At the Newtonian order, explicit $H^0_{+,\times}$ expressions can be extracted 
from Eqs.~(\ref{eq: hpx_qu}), and we list the higher order terms that appear
at $0.5$PN and 1PN orders in Appendix~\ref{sec: h_+,x}.
With the help of 1PN versions of the various relations derived in the previous 
subsection, we obtain 
\begin{align}
\label{eq:hxp1A}
	h_{+,\times} =&\; \frac{G\,m\,\eta}{R'\,c^2}x\sum_{p,q=0}^{\infty} \Big\{ \left[ a_{+,\times}^{p,q} \cos(pl)  + b_{+,\times}^{p,q} \sin(pl) \right]\cos(q\lambda) \nonumber\\
				&+\left[ c_{+,\times}^{p,q} \cos(pl)  + d_{+,\times}^{p,q} \sin(pl) \right]\sin(q\lambda)	\Big\}\,.
\end{align}
To show a glimpse of our final result, we display 
certain 1PN-accurate Fourier coefficients, truncated at ${\cal O}(e_t^3)$:
\begin{widetext}
	\begin{subequations}
	\label{eq:hxp1B}
		\begin{align}
			a^{0,1}_+ =&\; c_{1\beta}\,s_i\,\delta\,\sqrt{x}\left[-\frac{1}{4}\left(1+c_i^2\right)\left(1+2e_t^2\right) -1+2e_t^2\right] \,,\\
			a^{0,2}_+ =&\; c_{2\beta}\bigg\{ \left(1+c_i^2\right)\left(-2+5e_t^2\right) + x \bigg[ \frac{8}{3} \left(1-5e_t^2\right)\left(1-3\eta\right) \nonumber\\
						&+ \left(1+c_i^2\right)\left(3 +\frac{11\eta}{3} + \frac{1}{6}e_t^2 \left(315-151\eta\right) + 
						\frac{2}{3}s_i^2 \left(1+e_t^2\right)\left(1 - 3\eta\right)\right) \bigg] \bigg\} \,,\\
			a^{0,3}_+ =&\; \frac{9}{4}\,c_{3\beta}\,s_i\,\left(1+c_i^2\right)\delta\,\sqrt{x}\left(1-6e_t^2\right)  \,,\\
			a^{0,4}_+ =&\; -\frac{8}{3}\,c_{4\beta}\,s_i^2\,\left(1+c_i^2\right)x\left(1-3\eta\right)\left(1-11e_t^2\right) \,,\\
			c^{0,1}_+ =&\; s_{1\beta}\,s_i\,\delta\,\sqrt{x}\left[-\frac{1}{4}\left(1+c_i^2\right)\left(1+2e_t^2\right) -1+2e_t^2\right] \,,\\
			c^{0,2}_+ =&\; s_{2\beta}\bigg\{ \left(1+c_i^2\right)\left(-2+5e_t^2\right) + x \bigg[ \frac{8}{3} \left(1-5e_t^2\right)\left(1-3\eta\right)  \nonumber\\
						&+ \left(1+c_i^2\right)\left(3 +\frac{11\eta}{3} + \frac{1}{6}e_t^2 \left(315-151\eta\right) + 
						\frac{2}{3}s_i^2 \left(1+e_t^2\right)\left(1 - 3\eta\right)\right) \bigg] \bigg\} \,,\\
			c^{0,3}_+ =&\;  \frac{9}{4}\,s_{3\beta}\,s_i\,\left(1+c_i^2\right)\delta\,\sqrt{x}\left(1-6e_t^2\right) \,,  \\
			c^{0,4}_+ =&\; -\frac{8}{3}\,s_{4\beta}\,s_i^2\,\left(1+c_i^2\right)x\left(1-3\eta\right)\left(1-11e_t^2\right) \,,
		\end{align}
	\end{subequations}
\end{widetext}
where $c_{k\beta}$ and $s_{k\beta}$ stand for $\cos(k\beta)$ and $\sin(k\beta)$
and we list only those coefficients that survive in the circular limit.
We have verified that these coefficients are consistent with the 1PN-accurate 
amplitude-corrected $h_{+,\times}$ for quasicircular 
inspirals, provided in Ref.~\cite{BIWW}. 
This exercise demonstrates the ability of our inputs to compute analytic PN-accurate amplitude-corrected expressions
for $h_{+,\times}$ as a sum over {\it harmonics} in $l$.

Another important check of our approach is that we should also be able to 
reproduce Eqs.~(3.6)-(3.10) in Ref.~\cite{YABW} while restricting our attention 
to the quadrupolar order $h_{+,\times}$ from eccentric binaries in Newtonian eccentric orbits.
We use our Eq.~(\ref{eq: hpx_qu}) which provides the quadrupolar order 
$h_{+,\times}$ and the Newtonian version of our results from the previous subsection to obtain
\begin{align}
	h^0_{+,\times} &= -\frac{G\,m\,\eta}{c^2\,R'}x\sum_{p=0}^{\infty} \left[ \mathcal{C}_{+,\times}^p \cos(pl)  + \mathcal{S}_{+,\times}^p \sin(pl)\right] \,.
\end{align}
We list below $p=1$ coefficients accurate to ${\cal O}(e^8)$:
\begin{subequations}
	\begin{align}
		\label{eq: C+(1)}\mathcal{C}_+^1 =&\; s_i^2 \left(-e+\frac{e^3}{8}-\frac{e^5}{192} + \frac{e^7}{9216}\right) + c_{2\beta}(1+c_i^2) \nonumber\\
				&\times\left(-\frac{3e}{2} + \frac{2e^3}{3} -\frac{37e^5}{768} + \frac{11e^7}{7680}\right) \,,\\
		\mathcal{S}_+^1 =&\; s_{2\beta}(1+c_i^2) \nonumber\\
				&\times\left(-\frac{3e}{2} + \frac{23e^3}{24} + \frac{19e^5}{256} + \frac{371e^7}{5120}\right) \,,\\
		\mathcal{C}_\times^1 =&\; s_{2\beta} c_i \left(3e -\frac{4e^3}{3} + \frac{37e^5}{384} -\frac{11e^7}{3840} \right) \,,\\
		\mathcal{S}_\times^1 =&\; c_{2\beta}c_i \left(-3e+\frac{23e^3}{12} + \frac{19e^5}{128} + \frac{371e^7}{2560}\right) \,.
	\end{align}
\end{subequations}
Note that these are Newtonian order expressions and $e$ thus stands for the standard Newtonian eccentricity.
A close inspection reveals that our coefficients $\mathcal{S}_+^1$, $\mathcal{C}_\times^1$ and $\mathcal{S}_\times^1$ are identical to those 
given by Eqs.~(3.6)-(3.10) of Ref.~\cite{YABW}.
However, the coefficient of the $s_i^2$ term that appears in $\mathcal{C}_+^1$ is the negative of what is listed in Eq.~(3.7) of Ref.~\cite{YABW}.
To explore the origin of the above difference, we express our Eq.~(\ref{eq: hpx_qu}) in terms of the true anomaly (or the orbital phase)
with the help of the well-known classical Keplerian formulas
$(1-e\,\cos u) = (1-e^2)/( 1+ e\, \cos v)$, \quad$\sin u = (1- e^2)^{1/2}\,\sin v /(1+ e\, \cos v)$,
that connect true and eccentric anomaly.
The resulting expression for $h_{+}^{0} $ reads
\begin{widetext}
	\begin{align}
	\label{eq: hp_qv}
	h_{+}^{0} &= -\frac{ G\,m\, \eta}{c^2\,R'} \, \frac{x}{(1-e^2)} \left\{
	\left(1+c_i^2\right)\left( 2 \cos(2v-2\beta) + \frac{5e}{2} \cos(v-2\beta) + \frac{e}{2}\cos(3v-2\beta) + e^2 \cos(2\beta) \right)
	-s_i^2\left(e\cos v + e^2\right)\right\} \,.
	\end{align}
\end{widetext}
We observe that the above expression differs from Eq.~(3.1) of Ref.~\cite{YABW} in the sign of 
the $s_i^2$ term. This is indeed the reason why the sign of the $s_i^2$ term in our $\mathcal{C}_+^1$ 
differs from its counterpart, given in Eq. (3.7) of Ref.~\cite{YABW}.
In contrast, our Eq.~(\ref{eq: hp_qv}) is consistent with Eqs.~(30)-(32) of Ref.~\cite{Wahlquist87}.
Note that the relevant expressions of Ref.~\cite{Wahlquist87} are more general than ours. 
However, they can be compared to our Eq.~(\ref{eq: hp_qv}) 
by making the following substitutions:
$\theta\rightarrow v$, $\theta_{n}\rightarrow\beta$, $\phi\rightarrow 0$, $\theta_{p}\rightarrow 0$,
while using $\Phi=v-\beta$ at Newtonian order.
It turns out that the above-mentioned sign difference may
be associated with the convention adapted for defining
($\iota$, $\beta$) in the above calculations \cite{NY_private}.
At present, it is not very clear to us which convention is more appropriate 
while constructing GW response function from the amplitude corrected expressions for 
$h_+$ and $h_\times$.
The amplitude-corrected PN-accurate versions of these 
GW response functions will be reported elsewhere.

\section{A brief summary and possible extensions}\label{sec: conclusion}

We derived a compact and  elegant  solution to the 3PN-accurate Kepler 
equation, present in 
the generalized quasi-Keplerian parametrization for compact binaries 
in eccentric orbits.
This result crucially depends on certain 3PN-accurate infinite series expressions for trigonometric 
functions of $v$ in terms of $u$.
We probed the accuracy and correctness of our solution using analytical and numerical methods.
In Sec.~\ref{waveform_Fourier}, we provided PN-accurate crucial inputs that will be required to compute amplitude 
corrected GW polarization states as sum over harmonics in $l$.
The explicit use of these PN-accurate relations is demonstrated by computing 1PN-accurate analytic 
amplitude-corrected expressions for $h_{+,\times}(l)$. Detailed derivations of various PN-accurate 
relations are provided in the appendices.

It will be interesting to extend the present analysis for compact binaries in hyperbolic orbits.
This requires a 3PN-accurate Keplerian-type parametric solution for compact binaries 
in hyperbolic orbits and this is currently under investigation.
It will also be interesting to include spin effects into these computations with the help of Ref.~\cite{KJ10}.
Additionally, it will be worthwhile to compute 
fully analytic 3PN-accurate amplitude-corrected expressions for $h_{+,\times}$ with the help of 
our compact expressions and Ref.~\cite{MAI15}, that provides inputs to compute  amplitude-corrected 
$h_{+,\times}$ in terms of dynamical variables.

\acknowledgments
We thank Nico Yunes for informative discussions.
We thank Maria Haney for a first review and fruitful comments.
Y.~B. is supported by the Swiss National Science Foundation.
A.~G. would like to acknowledge the hospitality of the University of Zurich
during the initial stages of this collaboration. A.~K. acknowledges support from
the H2020-MSCA-RISE-2015 Grant No. StronGrHEP-690904. This work was supported 
by the Centre National d'{\'E}tudes Spatiales.

\appendix

\begin{widetext}

\section{Alternative solution to the PN-accurate Kepler Equation}\label{sec: alternative solution}

An alternative solution to the 3PN-accurate Kepler equation can be obtained in the following way. Rewrite Eq.~(\ref{eq: 3PN_KE}) as
\begin{align} \label{eq: 3PN_KE_alt}
	u-e_{t}\sin u &= l' \equiv l+\delta l \,,
\end{align}
where $\delta l$, a small perturbation to $l$, is given by
\begin{align}
	\delta l =&\; - \left(g_{4t} + g_{6t} \right)(v-u) - \left(f_{4t} + f_{6t} \right)\sin v  - i_{6t} \sin(2v) - h_{6t} \sin(3v)\,.
\end{align}
Eq.~(\ref{eq: 3PN_KE_alt}) looks like the classical Kepler equation, but with a mean anomaly $l'$. The solution to this equation can be written formally using Eq.~(\ref{eq: N_KE_Solution}) as
\begin{align}
	u=&\;l'+\sum_{k=1}^{\infty}\frac{2}{k}J_{k}(ke_t)\sin(kl')
\end{align}
Expanding in the small parameter $\delta l$,
\begin{align}\label{eq: KE_alternate_soln_step1}
	u=&\;l+\delta l+\sum_{k=1}^{\infty}\frac{2}{k}J_{k}(ke_{t})\sin(kl)+2\delta l\sum_{k=1}^{\infty}J_{k}(ke_{t})\cos(kl)	\nonumber \\
	=&\;l+\sum_{k=1}^{\infty}\frac{2}{k}J_{k}(ke_{t})\sin(kl)+\delta l\sum_{k=-\infty}^{\infty}J_{k}(ke_{t})\cos(kl)
\end{align}
Using Eqs.~(\ref{eq:3PN_relations}), we can write $\delta l$ as
\begin{alignat}{1}
\delta l & =-(f_{4t}+f_{6t})\frac{2\sqrt{1-e_{\phi}^{2}}}{e_{\phi}}\sum_{s=1}^{\infty}\frac{1}{s}\left(\sum_{j=1}^{\infty}\beta_{\phi}^{j}j\left[J_{s-j}(se_{t})+J_{s+j}(se_{t})\right]\right)\sin(sl)\nonumber \\
 & \quad-(g_{4t}+g_{6t})2\sum_{s=1}^{\infty}\frac{1}{s}\left(\sum_{j=1}^{\infty}\beta_{\phi}^{j}\left[J_{s-j}(se_{t})+J_{s+j}(se_{t})\right]\right)\sin(sl)\nonumber \\
 & \quad-i_{6t}\frac{4\sqrt{1-e_{\phi}^{2}}}{e_{\phi}^{2}}\sum_{s=1}^{\infty}\frac{1}{s}\left(\sum_{j=1}^{\infty}\beta_{\phi}^{j}j\left(j\sqrt{1-e_{\phi}^{2}}-1\right)\left[J_{s-j}(se_{t})+J_{s+j}(se_{t})\right]\right)\sin(sl)\nonumber \\
 & \quad-h_{6t}\frac{2\sqrt{1-e_{\phi}^{2}}}{e_{\phi}^{3}}\sum_{s=1}^{\infty}\frac{1}{s}\left(\sum_{j=1}^{\infty}\beta_{\phi}^{j}j\left(4-e_{\phi}^{2}-6j\sqrt{1-e_{\phi}^{2}}+2j^{2}(1-e_{\phi}^{2})\right)\left[J_{s-j}(se_{t})+J_{s+j}(se_{t})\right]\right)\sin(sl).
\end{alignat}
Invoking Eq.~(\ref{eq: alpha3PN}) for 
$\alpha_j$, we can rewrite 
\begin{align}
	\delta l=-\sum_{s=1}^{\infty}\sum_{j=1}^{\infty}\alpha_{j}\frac{j}{s}\left[J_{s-j}(se_{t})+J_{s+j}(se_{t})\right]\sin(sl) \,.
\end{align}
Substituting into Eq.~(\ref{eq: KE_alternate_soln_step1}), the above solution becomes
\begin{align}
	u=&\;l+\sum_{k=1}^{\infty}\frac{2}{k}J_{k}(ke_{t})\sin(kl) -\sum_{s=1}^{\infty}\sum_{j=1}^{\infty}\alpha_{j}\,\frac{j}{s}\left[J_{s-j}(se_{t})+J_{s+j}(se_{t})\right]\sin(sl) \sum_{k=-\infty}^{\infty}J_{k}(ke_{t})\cos(kl) \nonumber \\
	=&\;l+\sum_{k=1}^{\infty}\frac{2}{k}J_{k}(ke_{t})\sin(kl) -\sum_{s=1}^{\infty}\sum_{j=1}^{\infty}\alpha_{j}\frac{j}{s}\left[J_{s-j}(se_{t})+J_{s+j}(se_{t})\right]  \sum_{k=-\infty}^{\infty}J_{k}(ke_{t})\frac{1}{2}(\sin((k+s)l)-\sin((k-s)l)) \nonumber \\
	=&\;l+\sum_{k=1}^{\infty}\left[\frac{2}{k}J_{k}(ke_{t}) +\sum_{j=1}^{\infty}\alpha_{j}\sum_{s=1}^{\infty}\frac{j}{s}\left[J_{s-j}(se_{t})+J_{s+j}(se_{t})\right] \left[J_{k+s}((k+s)e_{t})-J_{k-s}((k-s)e_{t})\right]\right]\sin(kl) \,.
\end{align}
Therefore, we have $ u = l + \sum_{s=1}^{\infty}\,A_s\, \sin(sl)$ with 
\begin{align}
	A_{s} &= \frac{2}{s}J_{s}(se_{t})+\sum_{j=1}^{\infty}\alpha_{j}\sum_{k=1}^{\infty}\frac{j}{k}\left[J_{k-j}(ke_{t})+J_{k+j}(ke_{t})\right] \left[J_{s+k}((s+k)e_{t})-J_{s-k}((s-k)e_{t})\right] \,.
\end{align}
We have checked that this expression indeed matches with Eq.~(\ref{eq: KE solution}) 
when expanded to $\mathcal{O}(e_t^{40})$.

\section{Elegant series expansions for the required $v-u$ and $\sin(jv)$}\label{sec: trigonometric relations}

This appendix, as noted earlier, provides the derivation of Eqs.~(\ref{eq:3PN_relations}).
We begin by expressing the relation between the true  and eccentric anomaly as
\begin{align}
	\label{eq:app1}
	\tan\frac{v}{2}=\sqrt{\frac{1+e}{1-e}}\tan\frac{u}{2} \,,
\end{align}
where $e$ stands for the usual orbital eccentricity in the
Newtonian description or $e_{\phi}$ of the post-Newtonian approach.
Introduce  $\beta$ such that 
\begin{align}
	\frac{1+\beta}{1-\beta}=\sqrt{\frac{1+e}{1-e}}\,.
\end{align}
For eccentric binaries, it is convenient to express $\beta$ as 
$\frac{1-\sqrt{1-e^{2}}}{e}$.
This allows us to introduce the following 
popular series expansion for $v-u$ \cite{KE_Book}
\begin{align}
	v-u & = 2\sum_{n=1}^{\infty}\frac{\beta^{n}}{n}\sin(nu) \,.
\end{align}
We have verified that this series expansion is fully consistent 
with an exact relation for $v-u $, derived in Ref.~\cite{KG06}, namely
\begin{align}
	v - u  = 2 \tan^{-1}
	\left(
	\frac{ \beta \sin u }{ 1 - \beta \cos u }
	\right)\,.
\end{align}
The above series expansion for $v-u$ is indeed 
one of the series expansions required to tackle 
the PN-accurate Kepler equation. We are now in a position to derive similar compact series expansions for 
$\sin v$, $\sin(2v)$, $\sin(3v)$ etc.
The above relation connecting tangents of $v$ and $u$ 
may be written as
\begin{align}
\label{eq:tanv_2}
	\tan\frac{v}{2}=\frac{1+\beta}{1-\beta}\tan\frac{u}{2}\,.
\end{align}
Invoking the complex exponential representation of the tangent function, we 
write Eq.~(\ref{eq:tanv_2}) as
\begin{align}
	\frac{e^{-i\frac{v}{2}}-e^{i\frac{v}{2}}}{e^{-i\frac{v}{2}}+e^{i\frac{v}{2}}}=\frac{1+\beta}{1-\beta}\;\frac{e^{-i\frac{u}{2}}-e^{i\frac{u}{2}}}{e^{-i\frac{u}{2}}+e^{i\frac{u}{2}}}\,.
\end{align}
This leads to
\begin{align}
	e^{iv} & = \frac{e^{iu}-\beta}{1-\beta e^{iu}}\,.
\end{align}
Expanding this in powers of $e^{iu}$, we immediately get
\begin{align}
	e^{iv}	=-\beta+\frac{2\sqrt{1-e^{2}}}{e}\sum_{s=1}^{\infty}\beta^{s}e^{isu}\,.
\end{align}
Taking the imaginary part, we find
\begin{align}\label{eq:sinv}
	\sin(v)	=\frac{2\sqrt{1-e^{2}}}{e}\sum_{s=1}^{\infty}\beta^{s}\sin(su) \,.
\end{align}
For $\sin(2v)$, we can expand $e^{2iv}$ in a power series
\begin{align}
	e^{2iv}	&=\left( -\beta+\frac{2\sqrt{1-e^{2}}}{e}\sum_{s=1}^{\infty}\beta^{s}e^{isu} \right)^2 
	=\frac{\left(2-e^{2}\right)-2\sqrt{1-e^{2}}}{e^{2}}+\frac{4\sqrt{1-e^{2}}}{e^{2}}\sum_{s=1}^{\infty}\beta^{s}\left(s\sqrt{1-e^{2}}-1\right)e^{isu} \,.
\end{align}
This leads to 
\begin{align}\label{eq:sin2v}
	\sin(2v) =\frac{4\sqrt{1-e^{2}}}{e^{2}}\sum_{s=1}^{\infty}\beta^{s}\left(s\sqrt{1-e^{2}}-1\right)\sin(su) \,.
\end{align}

It is possible to check the correctness of these expressions by computing them with an 
independent method. In what follows, we briefly explain a different derivation of the
above $\sin(2v)$ expression.
This approach requires us to use the above-listed series expansion for $\sin v$ and the following expression
for $\cos v$, namely 
\begin{align}
	\cos v &= -\beta + 2\frac{\sqrt{1-e^2}}{e}\sum_{s=1}^{\infty}\beta^s \cos(su) \,.
\end{align} 
We use these series expansions for $\sin v$ and $\cos v$ to 
express $\sin(2v)$ as
\begin{align}\label{eq: sin2v_1}
	\sin(2v) &= 2 \sin v\cos v
	= 2\left(\frac{2 \sqrt{1-e^2}}{e} \sum_{i=1}^{\infty}\beta^i\sin(iu) \right) \left(-\beta + 2\frac{\sqrt{1-e^2}}{e}\sum_{j=1}^{\infty}\beta^j \cos(ju) \right) \nonumber\\
	&= -\frac{4 \sqrt{1-e^2}}{e} \sum_{i=1}^{\infty}\beta^{i+1}\sin(iu) + 8\frac{1-e^2}{e^2}\sum_{i,j\geq 1}\beta^{i+j} \sin(iu)\cos(ju) \,.
\end{align}
The double sum in the second part can be rewritten by invoking the Cauchy product formula \cite{Watson1922}:
\begin{align}
	\sum_{i,j\geq 1}\beta^{i+j} \sin(iu)\cos(ju) &= \sum_{k=1}^{\infty}\sum_{s=1}^{k}\beta^k \cos(su)\sin((k-s)u) \nonumber \\
	&= \frac{1}{2} \sum_{k=1}^{\infty}\beta^k\sum_{s=1}^{k}[\sin(ku)-\sin((2s-k)u)] \nonumber \\
	&= \frac{1}{2} \sum_{k=1}^{\infty}\beta^k (k-1)\sin(ku) \,.
\end{align}
With the help of this formula Eq.~(\ref{eq: sin2v_1}) becomes
\begin{align} \label{eq: sin2v}
	\sin(2v) &= \frac{4 \sqrt{1-e^2}}{e} \sum_{s=1}^{\infty} \left(\frac{\sqrt{1-e^2}}{e}(s-1) - \beta \right)\beta^s \sin(su) \nonumber \\
	&= \frac{4 \sqrt{1-e^2}}{e^2} \sum_{s=1}^{\infty} \beta^s \left(s\sqrt{1-e^2} -1\right) \sin(su) \,.
\end{align}
This is clearly identical to the earlier derived expression for $\sin(2v)$.

To obtain such elegant series expansions for higher order $\sin(jv)$, we introduce $\varsigma(z) = \left(\frac{1}{\beta}\frac{z-\beta^2}{1-z}\right)^j$. A close inspection reveals that
$e^{ijv}$ is identical to $\varsigma(\beta e^{iu})$. We now give the general Taylor series of $\varsigma(z)$. First note that
\begin{subequations}
	\begin{align}
		\frac{1}{(1-z)^k} &= \sum_{n=0}^{\infty}\frac{1}{n!}\frac{(k+n-1)!}{(k-1)!} z^n = \sum_{n=0}^{\infty} \binom{n+k-1}{n}z^n \,,\\
		(z-\beta^2)^j &= \sum_{k=0}^{j}\binom{j}{k}z^k(-\beta^2)^{j-k} \,.
	\end{align}
\end{subequations}
From this we find that
\begin{align}
	\varsigma(z) =&\; \frac{1}{\beta^j}\frac{(z-\beta^2)^j}{(1-z)^j} = \frac{1}{\beta^j} 
	\left(\sum_{k=0}^{j}\binom{j}{k}z^k(-\beta^2)^{j-k} \right) 
	\left( \sum_{s=0}^{\infty} \binom{s+j-1}{s}z^s  \right) \nonumber \\
	=&\; (-\beta)^j \sum_{n=0}^{\infty}\left(\sum_{s=0}^{n}\binom{s+j-1}{s}
	\binom{j}{n-s}(-1)^{s-n} \beta^{2(s-n)} \right)z^n \,.
\end{align}
We can give an explicit expression for the inner sum in terms of the hypergeometric function ${}_2F_1$ and find
\begin{subequations}\label{eq: e^ijv}
	\begin{align}
	e^{ijv} &= \varsigma(\beta e^{iu}) = \sum_{n=0}^{\infty} \mathcal{E}_n^j e^{inu} \,,\\
	\mathcal{E}_0^j	&= (-\beta)^j \,,\\
	\mathcal{E}_{n>0}^j &= \binom{n-1}{n-j} {}_2F_1(-j,n;n-j+1;\beta^2) \beta^{n-j} \,.
	\end{align}
\end{subequations}
Also note that the negative harmonics are simply given by $e^{-ijv} = \sum_{n=0}^{\infty} \mathcal{E}_n^j e^{-inu}$.
From this result the series expansions of $\sin(jv)$ and $\cos(jv)$ are easily extracted to be
\begin{subequations}
	\begin{align}
		\sin(jv) &= \sum_{n=1}^{\infty} \mathcal{E}_n^j \sin(nu) \,,\\
		\cos(jv) &= \sum_{n=0}^{\infty} \mathcal{E}_n^j \cos(nu) \,.	
	\end{align}
\end{subequations}
It should be noted that these derivations indeed provide 
elegant and compact expressions for $\sin v$, $\sin(2v) $ and $\sin(3v)$
that are crucial for computing semianalytic solution to our 3PN-accurate Kepler equation.
Explicitly, the first few expressions are
\begin{subequations}
	\begin{align}
		\sin v =&\; \frac{2\sqrt{1-e^{2}}}{e}\sum_{s=1}^{\infty}\beta^{s}\sin(su)\,,\\
		\sin(2v) =&\;\frac{4\sqrt{1-e^{2}}}{e^{2}}\sum_{s=1}^{\infty}\beta^{s}\left(s\sqrt{1-e^{2}}-1\right)\sin(su)\,,\\
		\sin(3v) =&\; \frac{2\sqrt{1-e^{2}}}{e^{3}}\sum_{s=1}^{\infty}\beta^{s}\left(2\left(1-e^{2}\right)s^{2}-6\sqrt{1-e^{2}}s+4-e^{2}\right)\sin(su)\,,\\
		\sin(4v) =&\; \frac{8\sqrt{1-e^{2}}}{3e^{4}}\sum_{s=1}^{\infty}\beta^{s}\left(\left(1-e^{2}\right)^{3/2}s^{3}-6\left(1-e^{2}\right)s^{2}
					+\left(1-e^{2}\right)^{1/2}\left(11-2e^{2}\right)s+3\left(e^{2}-2\right)\right)\sin(su)\,,\\
		\sin(5v) =&\; \frac{2\sqrt{1-e^{2}}}{3e^{5}}\sum_{s=1}^{\infty}\beta^{s}\Big(2\left(1-e^{2}\right)^{2}s^{4}-20\left(1-e^{2}\right)^{3/2}s^{3} \nonumber\\
					&+10\left(1-e^2\right)\left(7-e^2\right)s^{2}-20\sqrt{1-e^{2}}\left(5-2e^{2}\right)s+48-36e^2+3e^4\Big)\sin(su)\,.
	\end{align} 
\end{subequations}

\section{PN-accurate expression for $v$ in terms of $l$}\label{sec: true anomaly}

We begin by describing in detail how one obtains 
the series expansion for the true anomaly $v = 2 \arctan{\left( \sqrt{\frac{1+e}{1-e}}\tan{\frac{u}{2}} \right)}$
in terms of the mean anomaly $l$ for the Keplerian parametrization.
The definition of $v$ allows us to write 
\begin{align}
	v - l= \sum_{s = 1}^{\infty}B_s \sin(sl) \,,
\end{align}
where the Fourier  coefficients are given by
\begin{align}
	B_s &= \frac{2}{\pi} \int_{0}^{\pi} (v-l) \sin(s l) dl
	= \frac{2}{s\pi} \int_{0}^{\pi} \cos(s l) \frac{dv}{du}du
	= \frac{2}{s\pi} \int_{0}^{\pi} \cos(s l) \frac{\sqrt{1-e^2}}{1-e \cos u}du \,.
\end{align}
We invoke now a familiar expression, namely 
\begin{align}
	\frac{\sqrt{1-e^2}}{1-e \cos u} &= 1 + 2 \sum_{j=1}^{\infty} \beta^j \cos(j u) \,,
\end{align}
with $\beta = (1-\sqrt{1-e^2})/e$. This leads to
\begin{align}
\label{eq: Fourier v}
	B_s &= \frac{2}{s\pi} \int_{0}^{\pi} \cos(s l)du + 
	\frac{2}{s\pi} \sum_{j=1}^{\infty}\beta^j \int_{0}^{\pi} \left\{\cos(s l + j u) + \cos(s l - j u)\right\} du \nonumber \\
	&= \frac{2}{s} J_s(s e) + \frac{2}{s} \sum_{j=1}^{\infty}\beta^j \left\{J_{s+j}(se) + J_{s-j}(se)\right\} \,.
\end{align}
where in the last step we invoked the usual integral definitions of the Bessel functions of the first kind.
This gives us our desired result
\begin{align}
	v &= l + \sum_{s=1}^{\infty} \frac{2}{s} \Bigg( J_s(s e) +  \sum_{j=1}^{\infty}\beta^j \left\{J_{s+j}(se) + J_{s-j}(se)\right\} \Bigg) \sin(s l) \,.
\end{align}

In the PN-accurate generalized quasi-Keplerian description,
the true anomaly is related to the eccentric anomaly by
\begin{align}
	v = 2 \arctan \left(\sqrt{\frac{1+e_{\phi}}{1-e_{\phi}}} \tan\frac{u}{2} \right) \,.
\end{align}
We invoke a Fourier series expansion of the true anomaly in terms of the mean anomaly
\begin{align}\label{eq: v}
	v = l + \sum_{s=1}^{\infty}B_s \sin(sl)\,.
\end{align}
It is fairly straightforward to write down the following expression for 
the constant coefficients $B_s$ 
\begin{align} \label{eq:Bs}
	B_s =& \frac{2}{s} \Bigg( J_s(s e_t) + \sum_{j = 1}^{\infty} \beta_{\phi}^j \left\{J_{s+j}(s e_t) + J_{s-j}(se_t) \right\} \Bigg) \nonumber \\
	& + \sum_{j=1}^{\infty} \Bigg(\alpha_j \left\{J_{s+j}(s e_t) - J_{s-j}(s e_t) \right\} \nonumber \\
	& + \beta_{\phi}^j \sum_{i = 1}^{\infty} \alpha_i \left\{J_{s+j+i}(se_t) - J_{s+j-i}(se_t) + J_{s-j+i}(se_t) - J_{s-j-i}(se_t) \right\}  \Bigg) \,.
\end{align}

\section{Product of Fourier series} \label{sec: Product of Fourier series}
 In what follows, we derive compact expressions for 
certain products of Fourier sine and cosine series. Explicitly, we consider the products
\begin{subequations}
	\begin{align}
		\left(\sum_{s=1}^{\infty}A_s \cos(sl)\right)\left(\sum_{k=1}^{\infty}B_k \cos(kl)\right) &= \sum_{n=0}^{\infty}P_n^{CC} \cos(nl) \,,\\
		\left(\sum_{s=1}^{\infty}A_s \sin(sl)\right)\left(\sum_{k=1}^{\infty}B_k \sin(kl)\right) &= \sum_{n=0}^{\infty}P_n^{SS} \cos(nl) \,,\\
		\left(\sum_{s=1}^{\infty}A_s \cos(sl)\right)\left(\sum_{k=1}^{\infty}B_k \sin(kl)\right) &= \sum_{n=1}^{\infty}P_n^{CS} \sin(nl) \,,
	\end{align}
\end{subequations}  
that will be crucial to obtain analytic time-domain $h_{+,\times}(l)$. 
We show in detail the derivation of the first product in the above equations.
Multiplying out the product and using the angle sum identity for cosine we get
\begin{align}
	\left(\sum_{s=1}^{\infty}A_s \cos(sl)\right)\left(\sum_{k=1}^{\infty}B_k \cos(kl)\right) &= \frac{1}{2} \sum_{s=1}^{\infty}\sum_{k=1}^{\infty} A_s B_k 
		\left\{ \cos((s+k)l) + \cos((s-k)l) \right\}\,.
\end{align}
We note that the first cosine factor will only contribute to frequencies $n = s+k$, while the second factor will contribute at $n = |s-k|$. The zero mode only appears in the second factor for $s = k$. Thus we can write
\begin{align}
	\frac{1}{2} \sum_{s=1}^{\infty}\sum_{k=1}^{\infty} A_s B_k 	\left\{ \cos((s+k)l) + \cos((s-k)l) \right\} =&\; \frac{1}{2}\sum_{s=1}^{\infty} A_s B_s \nonumber\\
	&+ \frac{1}{2}\sum_{n=1}^{\infty} \left( \sum_{s=1}^{\infty}\sum_{k=1}^{\infty} A_s B_k \left\{ \delta_{s+k,n} + \delta_{s-k,n} + \delta_{k-s,n} \right\} \right) \cos(nl) \nonumber\\
	=&\; \frac{1}{2}\sum_{s=1}^{\infty} A_s B_s \nonumber\\
	&+ \frac{1}{2}\sum_{n=1}^{\infty} \left( \sum_{s=1}^{n-1} A_sB_{n-s} + \sum_{s=n+1}^{\infty}A_sB_{s-n} + \sum_{s=1}^{\infty}A_sB_{s+n}\right) \cos(nl) \,.
\end{align}
This allows us to write 
\begin{subequations}
	\begin{align}
		\left(\sum_{s=1}^{\infty}A_s \cos(sl)\right)\left(\sum_{k=1}^{\infty}B_k \cos(kl)\right) &= \sum_{n=0}^{\infty}P_n^{CC} \cos(nl) \,,\\
		P_0^{CC} &= \frac{1}{2}\sum_{s=1}^{\infty} A_s B_s \,,\\
		P_{n>0}^{CC} &= \frac{1}{2} \left( \sum_{s=1}^{n-1} A_sB_{n-s} + \sum_{s=n+1}^{\infty}A_sB_{s-n} + \sum_{s=1}^{\infty}A_sB_{s+n}\right) \,.
	\end{align}
\end{subequations}
The other products can be derived in a similar fashion and they read
\begin{subequations}
	\begin{align}
		\left(\sum_{s=1}^{\infty}A_s \sin(sl)\right)\left(\sum_{k=1}^{\infty}B_k \sin(kl)\right) &= \sum_{n=0}^{\infty}P_n^{SS} \cos(nl) \,\\
		P_0^{SS} &= \frac{1}{2}\sum_{s=1}^{\infty} A_s B_s \,,\\
		P_{n>0}^{SS} &= \frac{1}{2} \left( \sum_{s=n+1}^{\infty}A_sB_{s-n} + \sum_{s=1}^{\infty}A_sB_{s+n} - \sum_{s=1}^{n-1} A_sB_{n-s} \right) \,.
	\end{align}
\end{subequations}
\begin{subequations}
	\begin{align}
		\left(\sum_{s=1}^{\infty}A_s \cos(sl)\right)\left(\sum_{k=1}^{\infty}B_k \sin(kl)\right) &= \sum_{n=1}^{\infty}P_n^{CS} \sin(nl) \,\\
		P_n^{CS} &= \frac{1}{2} \left( \sum_{s=1}^{n-1} A_sB_{n-s} - \sum_{s=n+1}^{\infty}A_sB_{s-n} + \sum_{s=1}^{\infty}A_sB_{s+n}\right) \,.
	\end{align}
\end{subequations}

\section{Fourier series of $e^{imW}$}\label{sec: Fourier series of exp(imW)}

We rewrite the Fourier series for $W$, given by Eq.~(\ref{eq: Fourier series of W}), as 
\begin{align}
	W(l) &= (v-l) + \sum_{s=1}^{\infty} \omega_s \sin(sl)\,,
\end{align}
where $\omega_s$ is simply given by
\begin{align}
	\omega_s = \mathcal{W}_s - B_s = k B_s + (f_{4\phi}+f_{6\phi}) \sigma_s^{2v} + (g_{4\phi}+g_{6\phi}) \sigma_s^{3v} + i_{6\phi} \sigma_s^{4v} + h_{6\phi} \sigma_s^{5v} \,.
\end{align}
We isolate the $v-l$ part for the ease of calculation. The harmonics $e^{imW}$ can then be written as
\begin{align}
	e^{im W} &= e^{im (v-l)}e^{im\sum_{s=1}^{\infty}\omega_s\sin(sl)}\,.
\end{align}
The first part of this can be expanded as a Fourier series using the results in Eqs.~(\ref{eq: Fourier series of e^ijv})
\begin{align}
	e^{im (v-l)} &= e^{-iml}e^{imv} = e^{-iml}\sum_{s=-\infty}^{\infty}\epsilon_s^{mv}e^{isl} = \sum_{s=-\infty}^{\infty}\epsilon_{s+m}^{mv}e^{isl}\,.
\end{align}
The second part contains only PN-accurate quantities, so it can be expanded in $x$ up to 3PN order, resulting in
\begin{align}\label{eq: exp of W}
	e^{im\sum_{s=1}^{\infty}\omega_s\sin(sl)} =&\; 1 + im \sum_{s=1}^{\infty}\omega_s\sin(sl) 
	-\frac{m^2}{2} \left(\sum_{s=1}^{\infty}\omega_s\sin(sl)\right)^2 
	-\frac{i m^3}{6} \left(\sum_{s=1}^{\infty}\omega_s\sin(sl)\right)^3 \,.
\end{align}
We now use the results from Appendix~\ref{sec: Product of Fourier series}
to expand the products of the Fourier sine series. We immediately see
\begin{subequations}
	\begin{align}
	\left(\sum_{s=1}^{\infty}\omega_s\sin(sl)\right)^2 =&\; \sum_{n=0}^{\infty} C_n \cos(nl) \,, \\
	C_0 =&\; \frac{1}{2}\sum_{s=1}^{\infty} (\omega_s)^2 \,,\\
	C_{n>0} =&\; \frac{1}{2} \Bigg( \sum_{s=n+1}^{\infty}\omega_s\omega_{s-n} + \sum_{s=1}^{\infty}\omega_s\omega_{s+n} 
	- \sum_{s=1}^{n-1}\omega_s\omega_{n-s} \Bigg) \,.
	\end{align}
\end{subequations}
Using this result, the triple product can be written as
\begin{align}
	\left(\sum_{s=1}^{\infty}\omega_s\sin(sl)\right)^3 =&\; \left(\sum_{s=1}^{\infty}\omega_s\sin(sl)\right)^2\left(\sum_{s=1}^{\infty}\omega_s\sin(sl)\right)
	=\; \left(\sum_{n=0}^{\infty} C_n \cos(nl)\right)\left(\sum_{s=1}^{\infty}\omega_s\sin(sl)\right) \nonumber\\
	=&\; C_0 \sum_{s=1}^{\infty}\omega_s\sin(sl) + \left(\sum_{n=1}^{\infty} C_n \cos(nl)\right)\left(\sum_{s=1}^{\infty}\omega_s\sin(sl)\right) \,.
\end{align}
The product of a cosine and sine series can be expanded using the result in Appendix~\ref{sec: Product of Fourier series} and we find 
\begin{subequations}
	\begin{align}
		\left(\sum_{s=1}^{\infty}\omega_s\sin(sl)\right)^3 =&\; \sum_{n=1}^{\infty} D_n \sin(nl) \,, \\
		D_n =&\; C_0 \omega_n + \frac{1}{2} \Bigg( \sum_{s=1}^{n-1} C_s\omega_{n-s} - \sum_{s=n+1}^{\infty}C_s\omega_{s-n} 
		+ \sum_{s=1}^{\infty}C_s\omega_{s+n}\Bigg) \nonumber\\
		=&\; \frac{1}{2} \Bigg( \sum_{s=0}^{n-1} C_s\omega_{n-s} - \sum_{s=n+1}^{\infty}C_s\omega_{s-n} 
		+ \sum_{s=0}^{\infty}C_s\omega_{s+n}\Bigg)\,.
	\end{align}
\end{subequations}
Eq.~(\ref{eq: exp of W}) can thus be decomposed into a Fourier series as
\begin{align}
	e^{im\sum_{s=1}^{\infty}\omega_s\sin(sl)} =&\; 1 -\frac{m^2}{2}C_0 - \sum_{s=1}^{\infty}\frac{m^2}{2} C_s\cos(sl) + i\sum_{s=1}^{\infty}\left(m\omega_s-\frac{m^3}{6} D_s\right)\sin(sl) \,.
\end{align}
Converting the sine and cosine series to an exponential Fourier series
\begin{subequations}
	\begin{align}
		e^{im\sum_{s=1}^{\infty}\omega_s\sin(sl)} &= \sum_{s=-\infty}^{\infty} \Omega^m_s e^{isl} \,,\\
		\Omega^m_0 &= 1 -\frac{m^2}{2}C_0 \,,\\
		\Omega^m_{s\ne0} &= \frac{1}{2}\left(-\frac{m^2}{2} C_{|s|} + \textnormal{sgn}(s)\left(m\omega_{|s|}-\frac{m^3}{6} D_{|s|}\right) \right)\,.
	\end{align}
\end{subequations}
We now put all of this together and find the Fourier decomposition of the harmonics of $W$
\begin{align}
	e^{im W} &= e^{im (v-l)}e^{im\sum_{s=1}^{\infty}\omega_s\sin(sl)}\nonumber\\
	&= \left(\sum_{s=-\infty}^{\infty}\epsilon_{s+m}^{mv}e^{isl}\right) \left(\sum_{k=-\infty}^{\infty} \Omega^m_k e^{ikl}\right) \nonumber\\
	&= \sum_{s=-\infty}^{\infty}\sum_{k=-\infty}^{\infty} \epsilon_{s+m}^{mv} \Omega^m_k e^{i(s+k)l} \nonumber \\
	&= \sum_{n=-\infty}^{\infty} \mathcal{P}_n^{mW} e^{inl} \,,
\end{align}
where the constant Fourier coefficients $\mathcal{P}_n^{mW}$ are given by
\begin{align}\label{eq: Fourier coefficients e^imW}
	\mathcal{P}_n^{mW} &= \sum_{s=-\infty}^{\infty} \epsilon_{s+m}^{mv} \Omega^m_{n-s} \,.
\end{align}

\section{1PN accurate expressions for $h_+$ and $h_{\times}$}\label{sec: h_+,x}

Employing inputs from Refs.~\cite{Gopakumar2002,DGI,THG16}, the amplitude corrected 
1PN accurate expressions for $h_{+,\times}$ may be written as
\begin{align}\label{eq: h_+x expansion}
	h_{+,\times}=\frac{G\,m\,\eta}{R'\,c^2} x \left\{H_{+,\times}^{0}+x^{0.5}\,H_{+,\times}^{0.5}+x\,H_{+,\times}^{1}\right\} \,.
\end{align}
The explicit expressions for $H_{+,\times}^{i}$ are given by
{\small{}
\begin{subequations}
\begin{align}
	\label{eq: hp_Newtonian}
	H_+^0= 			&\; \frac{1}{(1-\chi)^{2}}\left(-2\left(c_{i}^{2}+1\right)\sqrt{1-e_{t}^{2}}\xi\sin(2\Phi)+
					\left(c_{i}^{2}+1\right)\cos(2\Phi)\left(\left(2e_{t}^{2}-
					\chi^{2}\right)+\chi-2\right)+s_{i}^{2}(1-\chi)\chi\right) \,,\\
	H_+^{0.5} =		&\; \frac{1}{(1-\chi)^{3}}\frac{\delta s_{i}}{4}\left\{\left(c_{i}^{2}+1\right)\sqrt{1-e_{t}^{2}}
					\left(6\chi^{2}-7\chi-8e_{t}^{2}+9\right)\cos(3\Phi)+2\left(c_{i}^{2}+1\right)\xi
					\left(\chi^{2}-2\chi-4e_{t}^{2}+5\right)\sin(3\Phi)\right.\nonumber \\
					& \left.+\sqrt{1-e_{t}^{2}}(1-\chi)\left(\left(6c_{i}^{2}-2\right)\chi-c_{i}^{2}-5\right)\cos(\Phi)
					+2\left(1-3c_{i}^{2}\right)(1-\chi)^{2}\xi\sin(\Phi)\right\} \,,\\
	H_+^1 = 		&\; \frac{1}{(1-\chi)^{4}}\frac{1}{24}\left\{ 6(1-3\eta)\xi\sqrt{1-e_{t}^{2}}\left(c_{i}^{2}+1\right)
					\left(-4\chi^{2}+9\chi+8e_{t}^{2}-13\right)s_{i}^{2}\sin(4\Phi)\right.\nonumber \\
					& +(1-3\eta)\left(c_{i}^{2}+1\right)\left(-6\chi^{4}+18\chi^{3}+\left(48e_{t}^{2}-61\right)\chi^{2}+
					\left(65-69e_{t}^{2}\right)\chi-48e_{t}^{4}+117e_{t}^{2}-64\right)s_{i}^{2}\cos(4\Phi)\nonumber \\
					& +\frac{(1-\chi)}{\sqrt{1-e_{t}^{2}}}4\xi\left[\left((15-45\eta)e_{t}^{2}+45\eta+
					\chi\left((36\eta-12)e_{t}^{2}-36\eta+12\right)-15\right)c_{i}^{4}\right.\nonumber \\
					& +\left((20\eta+30)e_{t}^{2}-32\eta+\chi\left((-26\eta-6)
					e_{t}^{2}+38\eta-30\right)+6\right)c_{i}^{2}\nonumber \\
					& \left.+(39-7\eta)e_{t}^{2}-5\eta+\chi\left((10\eta-18)
					e_{t}^{2}+2\eta-18\right)-3\right]\sin(2\Phi)\nonumber \\
					& +\frac{(1-\chi)}{\left(1-e_{t}^{2}\right)}4\left[\left((27\eta-9)e_{t}^{4}+(13-39\eta)e_{t}^{2}+12\eta+
					\chi^{3}\left((18\eta-6)e_{t}^{2}-18\eta+6\right)+\chi^{2}
					\left((12-36\eta)e_{t}^{2}+36\eta-12\right)\right.\right.\nonumber \\
					& \left.+\chi\left((12-36\eta)e_{t}^{4}+(75\eta-25)e_{t}^{2}-39\eta+13\right)-4\right)c_{i}^{4}\nonumber \\
					& +\left((34\eta-48)e_{t}^{4}+(30-56\eta)e_{t}^{2}+22\eta+\chi^{2}\left((26\eta+6)e_{t}^{2}-
					26\eta+18\right)+\chi^{3}\left((-13\eta-3)e_{t}^{2}+13\eta-9\right)\right.\nonumber \\
					& \left.+\chi\left((26\eta+6)e_{t}^{4}+(51-77\eta)e_{t}^{2}+51\eta-69\right)+18\right)c_{i}^{2}\nonumber \\
					& +(-11\eta-33)e_{t}^{4}+(49\eta-5)e_{t}^{2}-38\eta+\chi^{3}\left((5\eta-9)e_{t}^{2}-5\eta-3\right)\nonumber \\
					& \left.+\chi^{2}\left((18-10\eta)e_{t}^{2}+10\eta+6\right)+\chi\left((18-10\eta)e_{t}^{4}+
					(26-2\eta)e_{t}^{2}+12\eta-56\right)+38\right]\cos(2\Phi)\nonumber \\
					& +\frac{(1-\chi)}{\left(1-e_{t}^{2}\right)}\left[\left((15-45\eta)e_{t}^{4}+(45\eta-15)e_{t}^{2}+
					\chi^{2}\left((108\eta-36)e_{t}^{2}-108\eta+36\right)+\chi\left((3-9\eta)
					e_{t}^{2}+9\eta-3\right)\right.\right.\nonumber \\
					& \left.+\chi^{3}\left((18-54\eta)e_{t}^{2}+54\eta-18\right)\right)c_{i}^{4}+\left((48\eta+72)e_{t}^{4}+
					(-48\eta-72)e_{t}^{2}+\chi\left((4\eta-60)e_{t}^{2}-4\eta+108\right)\right.\nonumber \\
					& \left.+\chi^{3}\left((52\eta+12)e_{t}^{2}-52\eta+36\right)+\chi^{2}\left((-104\eta-24)e_{t}^{2}+
					104\eta-72\right)\right)c_{i}^{2} +(-3\eta-87)e_{t}^{4}+(3\eta+87)e_{t}^{2} \nonumber \\
					&\left.\left.+\chi^{3}\left((2\eta-30)e_{t}^{2}-2\eta-18\right)+
					\chi\left((5\eta+57)e_{t}^{2}-5\eta-105\right)+\chi^{2}\left((60-4\eta)e_{t}^{2}+
					4\eta+36\right)\right]\frac{\text{}}{}\right\} \,,\\
	\label{eq: hx_Newtonian}
	H_\times^0 =	&\; \frac{1}{(1-\chi)^{2}}2c_{i}\left(2\sqrt{1-e_{t}^{2}}\xi\cos(2\Phi)+
					\left(2e_{t}^{2}-\chi^{2}+\chi-2\right)\sin(2\Phi)\right) \,,\\
	H_\times^{0.5} =&\; \frac{1}{(1-\chi)^{3}}\frac{\delta}{2}c_{i}s_{i}\left\{ 2\xi\left(-\chi^{2}+
					2\chi+4e_{t}^{2}-5\right)\cos(3\Phi)+\sqrt{1-e_{t}^{2}}\left(\left(6\chi^{2}-
					8e_{t}^{2}\right)-7\chi+9\right)\sin(3\Phi)\right.\nonumber \\
					& \left.+\sqrt{1-e_{t}^{2}}\left(1-\chi\right)\left(2\chi-3\right)\sin(\Phi)+
					2\xi\left(1-\chi\right)^{2}\cos(\Phi)\right\} \,,\\
	H_\times^1 = 	&\; \frac{1}{(1-\chi)^{4}}\frac{1}{12\left(1-e_{t}^{2}\right)}\left\{ \frac{}{}(1-3\eta)c_{i}
					\left(1-e_{t}^{2}\right)s_{i}^{2}\left(\chi^{2}\left(48e_{t}^{2}-61\right)+\chi\left(65-69e_{t}^{2}\right)-
					48e_{t}^{4}+117e_{t}^{2}-6\chi^{4}+18\chi^{3}-64\right)\sin(4\Phi)\right.\nonumber \\
					& +6(1-3\eta)\left(1-e_{t}^{2}\right){}^{3/2}c_{i}s_{i}^{2}\xi\left(4\chi^{2}-
					9\chi-8e_{t}^{2}+13\right)\cos(4\Phi)\nonumber \\
					& +2c_{i}\left[e_{t}^{4}\left(50\eta+\chi^{2}\left(20\eta+(12-36\eta)s_{i}^{2}-36\right)+
					\chi\left(-70\eta+(99\eta-33)s_{i}^{2}+126\right)+(21-63\eta)s_{i}^{2}-90\right)\right.\nonumber \\
					& -46\eta e_{t}^{2}+\chi^{4}e_{t}^{2}\left(-10\eta+(18\eta-6)s_{i}^{2}+18\right)+
					\chi^{3}e_{t}^{2}\left(30\eta+(18-54\eta)s_{i}^{2}-54\right)+\chi^{2}e_{t}^{2}
					\left(-16\eta+(111\eta-37)s_{i}^{2}-16\right)\nonumber \\
					& +\chi e_{t}^{2}\left(42\eta+(62-186\eta)s_{i}^{2}+14\right)+(111\eta-37)s_{i}^{2}e_{t}^{2}+
					38e_{t}^{2}-4\eta+\chi^{4}\left(10\eta+(6-18\eta)s_{i}^{2}+6\right) \nonumber\\
					&+\chi^{3}\left(-30\eta+(54\eta-18)s_{i}^{2}-18\right)+
					\chi^{2}\left(-4\eta+(25-75\eta)s_{i}^{2}+124\right)+
					\chi\left(28\eta+(87\eta-29)s_{i}^{2}-164\right) \nonumber\\
					&\left.+(16-48\eta)s_{i}^{2}+52\right]\sin(2\Phi) +4(1-\chi)c_{i}\sqrt{1-e_{t}^{2}}\xi\left[e_{t}^{2}\left(16\eta+
					\chi\left(-10\eta+(18\eta-6)s_{i}^{2}+18\right)+(9-27\eta)s_{i}^{2}-42\right)-4\eta\right.\nonumber \\
					& +\left.\chi\left(-2\eta+(6-18\eta)s_{i}^{2}+18\right)+(27\eta-9)s_{i}^{2}+6\right]\cos(2\Phi)
					-\left.6(3\eta-1)(1-\chi)c_{i}\xi\left(1-e_{t}^{2}\right)^{3/2}s_{i}^{2}\right\} \,,
\end{align}
\end{subequations}
}
where $c_i=\cos(\iota)$, $s_i=\sin(\iota)$, $\chi=e_t \cos u$, $\xi=e_t \sin u$, $\Phi=\phi-\beta$ and $\delta=(m_1-m_2)/m$.
The above expressions, as noted earlier, are required to compute fully analytic $h_{+,\times}(l)$, given Eqs.~(\ref{eq:hxp1A}) and (\ref{eq:hxp1B}).

\end{widetext}

\bibliographystyle{apsrev4-1}
\bibliography{ke-paper}

\end{document}